\documentclass[onecolumn]{aastex61}
\pdfoutput=1
\usepackage{amsmath}
\usepackage{verbatim}
\usepackage{graphicx}	
\usepackage{booktabs}
\usepackage{color}
\usepackage{lmodern}
\usepackage{epstopdf}
\usepackage{physics}

\newcommand{\ix}[2]{#1_{\mathrm{#2}}}

\renewcommand{\vec}[1]{ \mbox{\boldmath $#1$} }

\received{October 9, 2017}
\revised{April 12, 2018}
\accepted{April 23, 2018}

\submitjournal{AJ}

\shorttitle{Fast and accurate radiative transfer}
\shortauthors{Stamer \& Inutsuka}

\begin{document}

\title{A fast and accurate method of radiation hydrodynamics calculation in spherical symmetry}

\author{Torsten Stamer}
\affil{Nagoya University \\
Furo-cho, Chikusa-ku, Nagoya, Japan}
\email{stamer.torsten@a.mbox.nagoya-u.ac.jp}

\author{Shu-ichiro Inutsuka}
\affil{Nagoya University \\
Furo-cho, Chikusa-ku, Nagoya, Japan}
\email{inutsuka@nagoya-u.jp}

\begin{abstract}

We develop a new numerical scheme for solving the radiative transfer equation in a spherically symmetric system. This scheme does not rely on any kind of diffusion approximation and it is accurate for optically thin, thick, and intermediate systems. In the limit of a homogeneously distributed extinction coefficient, our method is very accurate and exceptionally fast. We combine this fast method with a slower, but more generally applicable method to describe realistic problems. We perform various test calculations including a simplified protostellar collapse simulation. We also discuss possible future improvements.

\end{abstract}


\section{Introduction}

Theoretical investigations of many astrophysical problems require an accurate treatment of radiative transfer. Since analytic solutions of the transfer equation are generally unavailable except in trivial systems, numerical methods are employed. However, the complexity of the problem typically means that a realistic treatment is computationally expensive. One example of such an expensive problem is the simulation of protostellar collapse. The term refers to the process by which a molecular cloud core contracts under its own gravity to form a star, investigated in a seminal paper by \citet{Larson1969} and by others such as \citet{Winkler1980}, \citet{inu98} and \citet{Vaytet2017}. This process is initially isothermal: The cloud is optically thin, and so any heat gained through the compression of the gas is efficiently radiated away. As the collapse proceeds, the compressional heating increases and eventually catches up with the radiative cooling, so the temperature begins to rise. Depending on the initial conditions and opacity, this may happen before or after the system has become optically thick. The later evolution in the high-density region is essentially adiabatic. When the thermal pressure becomes large enough to stop the gravitational collapse, a hydrostatic object with a typical radius of a few AU forms. This is called the first hydrostatic core or the first Larson's core. The first core continues to accrete material from the surrounding envelope, and temperature continues to increase until, at about 2000 K, molecular hydrogen starts to dissociate. This is a strong cooling effect which reduces the pressure and triggers a second collapse, which only ends when all the hydrogen molecules have dissociated and a second hydrostatic core, much smaller and hotter than the first, has formed. This second core is also called a protostar; it continues to accrete matter and to heat up until nuclear fusion is ignited in its core.

Numerical simulations of protostellar collapse are challenging, which is in part due to the extremely wide range of size and density: When forming a solar-type star, typical values for the initial cloud core are $10^4 \mathrm {AU}$  and $10^{-19} \mathrm {g \: cm^{-3}}$, while the protostar is $10^{-3} \mathrm {AU}$ and $10^{-1} \mathrm{g \: cm^{-3}}$. Radiative transfer presents another difficulty: While the early, isothermal phase can be approximated by simple optically thin cooling, and the optically thick regions in the later stage by radiative diffusion, neither of these approximations is valid in the optically intermediate region near the photosphere. However, this is the most important region for the exchange with the outer medium, which determines the temperature structure and ultimately the entropy of the protostar. The latter determines the initial location in the HR-diagram and the subsequent evolution.

In recent years, numerical simulations have become more and more sophisticated. The number of physical processes included keeps increasing, among them magnetic field effects (\citet{Tomisaka2002, Machida2006, Commercon2010,Federrath2015}), protoplanetary disc formation (\citet{inu2010,Stamatellos2008,Machida2011,Tomida2015,Seifried2016,Gonzalez2015,Nordlund2014}), chemical evolution (\citet{Visser2015,Dzyurkevich2016,Hincelin2016}), and non-ideal MHD (\citet{Tsukamoto2015a, Tsukamoto2015b,Wurster2016,Masson2016}). There are now also synthetic observations comparable with actual data (\citet{Commercon2012a,Frimann2016,Seifried2016}). 

At the same time, much more simplified, spherically symmetric RHD simulations such as those by \citet{inu98,inu2000} and, more recently, \citet{Vaytet2011,Vaytet2012,Vaytet2013,Vaytet2017}, are still valuable: This one-dimensional modeling is still the only method to describe the evolution of the entire process of protostellar collapse, since multi-dimensional modeling has to mask the central region that requires extremely short timesteps. In this context, we aim to develop a radiative hydrodynamics scheme for spherical systems that is both fast and accurate, so that it can be used for a detailed exploration of the parameter space to investigate the effects of the initial conditions on properties such as protostellar mass, radius, and entropy. Note also that the present method can be combined with multi-dimensional modeling. In addition, even though collapse simulations are the primary motivation for this work, the method is not limited to this and may well be used in other systems that can be approximated as spherical.

Radiative transfer is often treated using either the flux-limited diffusion approximation, for example in works such as \citet{Bodenheimer1990} and \citet{Yorke1993}, or a method based on solving the moment equations of radiation (\citet{Castor1972,Buchler1979}). The latter requires a closure relation, which commonly appears in the form of the Eddington approximation, which assumes that the Eddington facor f, the ratio between the second and the zeroth moment of radiation, is equal to 1/3. This is exactly correct only in the case of isotropic radiation. An improvement upon this is the variable Eddington factor method (\citet{Tscharnuter1979}), in which the Eddington factor is no longer constant, but instead is calculated depending on the degree of anisotropy of the radiation field. Details of the numerical implementation can be found for instance in \citet{Mihalas1984} or \citet{Stone1992}, the basis for the ``ZEUS'' RHD-code.

Both the diffusion and Eddington approximation are very fast, but also somewhat crude, since they are only exact in optically thin / isotropic regions. On the other hand, the variable Eddington factor method is very accurate, but requires considerable computational effort. In this work, we develop a method that aims to combine the advantages of both. Our approach is different in the sense that it uses neither the diffusion approximation nor the moment equations. Instead, it is based on the radiative exchange between pairs of spherically symmetric shells. It does have some resemblance to ray-tracing schemes, especially in the case of the numerical angle integration discussed in section \ref{sec:Num}. 

This paper is structured in the following way: In section 2, we derive the basic equations on which our method is based, followed by a description of the numerical implementation. Section 3 presents the results of various test calculations, both for the case of a homogeneous extinction coefficient and for the more realistic, inhomogeneous case. Section 4 summarizes our work and presents ideas for possible future improvements.

\section{Method}
\subsection{Aims}
We develop a method to solve the radiation transport in a spherically symmetric fluid system. The aim is to calculate the rate of energy gain or loss due to radiation at any radius, i.e. ${dU / dt} \left(r\right)$, where $U$ is the specific internal energy of the fluid. Therefore the unit of this ``radiation term'', as we shall call it from now on, is power per mass.

Ultimately, our method is intended to couple to a hydrodynamic simulation by including the radiation term in the hydrodynamic energy equation. To first order, this can be done by subdividing the system into a finite number of shells, calculating the radiation term at some radius representative of each shell (such as the shell center), and assuming that it does not vary over the shell. For higher order calculations, one may interpolate the value of the radiation term between neighboring shells.

The radiation term is the sum of radiative heating and cooling. The latter is trivial even in a 3-D system since it depends only on the local values of density, opacity and temperature at position \vec{r}. But heating is far more complicated, as it depends on the local radiation field, which in turn is determined by the distribution of these variables across the entire system.

\subsection{Derivation of the basic equations}
\subsubsection{General derivation}
The monochromatic radiation transfer equation in three dimensions is

\begin{equation}
\frac{1}{c} \frac{dI_{\nu}}{dt} + \vec{n} \cdot \nabla I_{\nu} = \kappa_{\nu} \rho \left(S_{\nu} - I_{\nu} \right),
\label{eq:transeq}
\end{equation}
where $I_{\nu}$, $\kappa_{\nu}$ and $S_{\nu}$ are the intensity, opacity and source function at frequency $\nu$. In the following, we will leave out the index $\mathrm {\nu}$ for convenience.

Equation (\ref{eq:transeq}) combines the effects of emission and absorption of radiation by the material. We shall neglect the explicitly time dependent term $dI_{\nu}/dt$ throughout this paper. This is equivalent to assuming that the radiation field \textit {instantaneously} adjusts to any changes in the hydrodynamic variables. This assumption is valid as long as the time for light to cross the system is much shorter than the time for the hydrodynamic variables to change significantly, which is the case in most astrophysical systems.

With the above assumption, the solution is

\begin{align}
\begin{split}
I \left(\vec r, \vec n \right)&=\int_0^{D_{\mathrm B}} \kappa\left(\vec r + \vec n D \right) \rho\left(\vec r + \vec n D \right) S\left(\vec r + \vec n D \right)  e^{-\tau\left(\vec r, \vec r + \vec n D \right)} dD \\
&+ S_{\mathrm O} e^{-\tau_{\mathrm B}},
\label{eq:transeqsol}
\end{split}
\end{align}
where $\tau \left(\vec r, \vec r + \vec n D \right)$ is the optical depth between positions $\vec r$ and $\vec r + \vec n D$, while $D_{\mathrm B}$ and $\tau_{\mathrm B}$ refer to the distance and optical depth between $\vec r$ and the outer boundary of the system in direction $\vec n$. The first term represents radiation emitted by other points within the system, while the second term represents radiation coming from outside. The latter's intensity is determined by $S_{\mathrm O}$, the source function of the external radiation, which must be supplied as a boundary condition.

Equation (\ref{eq:transeqsol}) determines the direction-dependent intensity. The integral over all directions gives us the radiation energy density:

\begin{equation}
\epsilon \left(\vec r\right) = \frac{1}{c} \int_\Omega \left( \int_0^{D_{\mathrm B}} \kappa' \rho' S'  e^{-\tau} dD + S_{\mathrm O} e^{-\tau_{\mathrm B}} \right) d\Omega.
\label{eq:Erad}
\end{equation}
A prime on a position-dependent quantity indicates that it is to be evaluated at $\vec r + \vec n D$ rather than $\vec r$. 

Let us now consider the total radiation term (heating plus cooling). Assuming that each fluid element emits isotropically,

\begin{equation}
\frac{dU}{dt} = \kappa \left(c \epsilon - 4 \pi S\right),
\label{eq:RadTermSimple}
\end{equation}
and inserting the previously derived expression for $\epsilon$, we obtain

\begin{equation}
\frac{dU}{dt} = \kappa \left(\int_\Omega \left( \int_0^{D_{\mathrm B}} \kappa' \rho' S'  e^{-\tau} dD + S_O e^{-\tau_{\mathrm B}} \right) d\Omega - 4 \pi S\right).
\end{equation}
Assume a case where the temperature, and therefore the source function, is constant everywhere $\left(S=S'=S_{\mathrm O}\right)$. In this case, $\frac{dU}{dt}$ must be zero and it follows that

\begin{equation}
\int_\Omega \left( \int_0^{D_{\mathrm B}} \kappa' \rho' e^{-\tau} dD + e^{-\tau_{\mathrm B}} \right) d\Omega = 4\pi.
\label{eq:RadTermComplex}
\end{equation}
We can now replace the $4 \pi$ in equation (\ref{eq:RadTermSimple}) with this expression:

\begin{equation}
\frac{dU}{dt} = \kappa \int_\Omega \left( \int_0^{D_{\mathrm B}} \kappa' \rho' \left(S'-S\right)  e^{-\tau} dD + \left(S_{\mathrm O}-S\right) e^{-\tau_{\mathrm B}} \right) d\Omega.
\label{eq:BasicEqOrig}
\end{equation}

Consider the physical meaning of the integration in the first term: For all directions, we integrate outwards from a point $\vec r$ to the system's boundary. In other words, we are integrating over the volume of the physical system, but using a coordinate system whose origin lies at $\vec r$, so the volume element is $d\vec r' = D^2 d\Omega dD$, and we may write

\begin{equation}
\frac{dU}{dt} = \kappa \left( \int_V \kappa' \rho' \left(S'-S\right)  \frac{e^{-\tau}}{D^2} d\vec{r'} + \int_\Omega \left(S_{\mathrm O}-S\right) e^{-\tau_{\mathrm B}} d\Omega \right).
\label{eq:BasicEq}
\end{equation} 
As long as the integral is over the whole volume, we may use the usual coordinate system and volume element. From this point onward, we shall abbreviate the product $\kappa \rho$ as $\chi$ (the extinction/emission coefficient). The physical interpretation of this equation is as follows:

The volume integral represents the exchange of radiative energy between different positions within the system. Thus, the term with S (cooling term) is the radiation emitted at $\vec r$ and absorbed at some other point $\vec {r'}$, while the term with S' (heating term) is radiation emitted at $\vec {r'}$ and absorbed at $\vec r$.

Likewise, the $S_{\mathrm O}$-term in the solid angle integral represents radiation coming from outside the system which is absorbed at $\vec r$, while the $S$-term represents radiation emitted at $\vec {r}$ which reaches the boundary and escapes from the system.

The factors $\kappa \chi' e^{-\tau}/D^2$ and $\kappa \int_{\Omega} e^{-\tau_{\mathrm B}} d\Omega$ can be regarded as exchange coefficients, quantifying the efficiency of the radiative exchange between $\vec r$ and $\vec {r'}$ and $\vec r$ and the outside radiation field, respectively. Furthermore, equation (\ref{eq:RadTermComplex}) tells us that the volume integral over all of these exchange coefficients, plus the exchange coefficient for the boundary, must be equal to $4 \pi \kappa$. This is simply another way of saying that the radiative cooling is equal to $-4 \pi \kappa S$.

\subsubsection{Spherical Symmetry}
In principle, equation (\ref{eq:BasicEq}) can be integrated numerically to obtain the radiation term, but in realistic problems the computational cost is prohibitively high.

Many hydrodynamical problems in astrophysics can be reasonably well approximated as spherically symmetric, so that all variables depend only on radius. While this reduces the computational cost, it is still quite slow because the equations include the distance and optical depth between different points. Even in spherical symmetry, these depend on all 3 coordinates. In this section, we shall manipulate the equation in such a way as to reduce this computational load for the case of a spherically symmetric system.

\begin{figure}
		\plotone {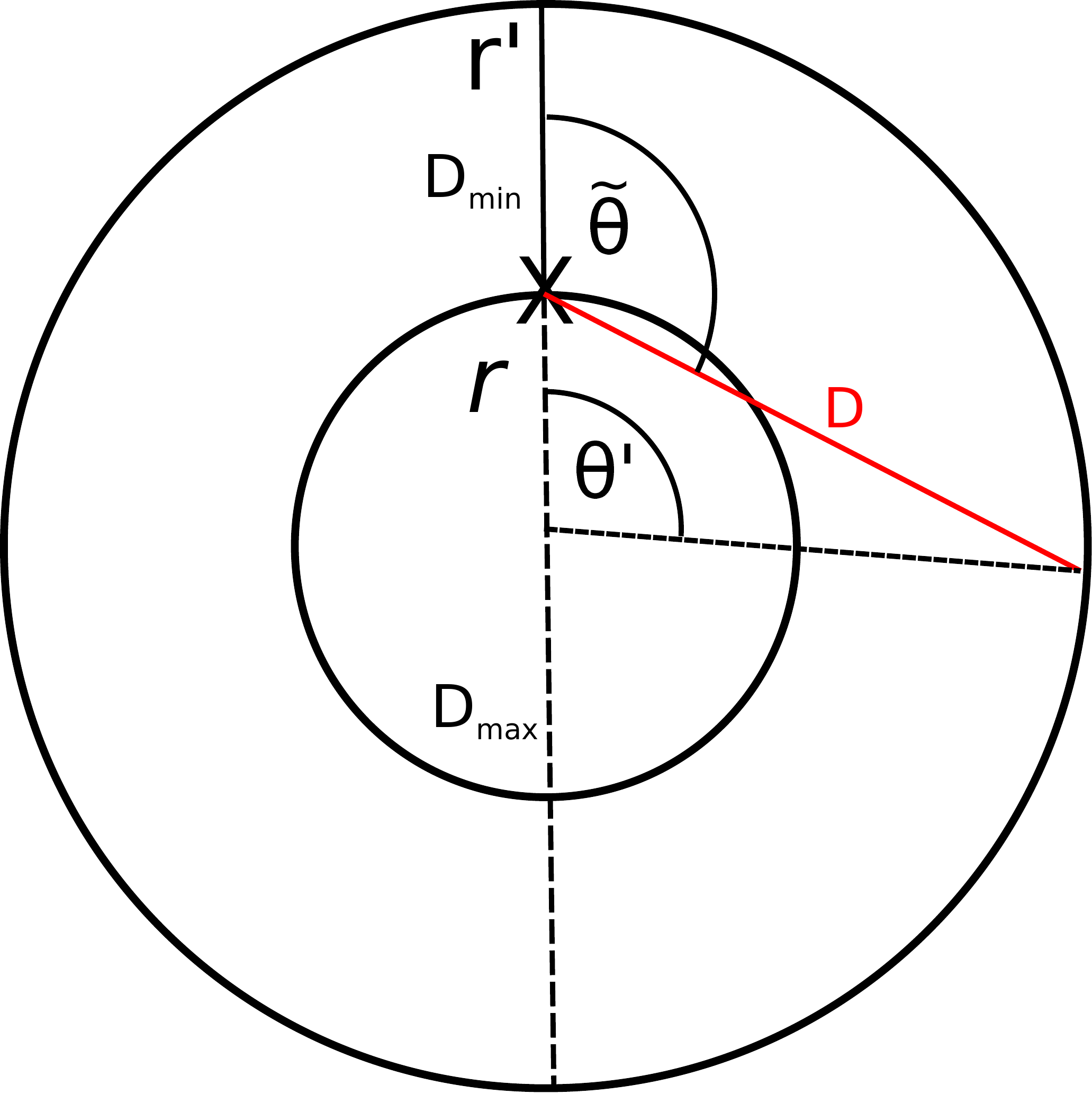}
	\caption{A schematic illustration of the situation in spherical symmetry, for the case $r'>r$. The idea is to calculate the radiative exchange between the point $\vec r$ and the infinitesimal shell r'. The red line represents a single ray between $\vec r$ and a point on r'. The full calculation requires an integration over all such rays. \newline
The picture shows the meaning of a number of important variables: The distance between the points D, the polar angle $\theta'$, the direction angle $\tilde \theta$, and the minimum and maximum distances between shells r and r', which occur at $\tilde \theta = \theta' = 0$ and $\tilde \theta = \theta' = \pi$, respectively. The optical depth $\tau$ is proportional to the distance, but unless $\chi$ is constant throughout the system, the proportionality constant depends on the direction of the ray.}
	\label{fig:SpherSym}
\end{figure}

Consider the radiative exchange between a \textit{point} $\vec r$ and an infinitesimally thin, spherical shell at radius r' (Figure \ref{fig:SpherSym}). Calculating the radiation term at point $\vec r$ is sufficient, because spherical symmetry implies that it is the same for every point with radius r.

If we choose the coordinate system such that $\vec r$ lies on the cartesian y-axis, there is no longer any dependency on the azimuthal angle, essentially reducing the system to 2-D. Equation (\ref{eq:BasicEq}) tells us that the contribution of the (infinitesimal) shell r' to the radiation term at $\vec r$ is

\begin{align}
\begin{split}
R_{\mathrm r \leftarrow \mathrm r'}&=\kappa \chi' \left(S'-S\right) \int_{\theta'} \int_{\phi'} \frac{e^{-\tau}}{D^2} {r'}^2 \sin\left(\theta'\right) d\theta' d\phi' dr' \\
&=2\pi \kappa \chi' \left(S'-S\right) \int_{\theta'} \frac{e^{-\tau}}{D^2} {r'}^2 \sin\left(\theta'\right) d\theta' dr' \\
&=2\pi \kappa \chi' \left(S'-S\right) A_{\mathrm r \leftarrow \mathrm r'}. 
\label{eq:BasicEqDiff}
\end{split}
\end{align}
Numerical integration over r' is of course unavoidable. The main challenge is the calculation of the angle integral: 
\begin{equation}
A_{\mathrm r \leftarrow \mathrm r'}={r'}^2 \int_{0}^{\pi} \frac{e^{-\tau\left(\theta'\right)}}{D^2\left(\theta'\right)} \sin\left(\theta'\right) d\theta' dr'
\end{equation}

Figure \ref{fig:SpherSym} illustrates the physical meaning of the variables in this equation. The next section deals with methods of calculating $A_{\mathrm r \leftarrow \mathrm r'}$.

\subsection{Numerical Implementation}
\label{sec:NumImp}
\subsubsection{General Considerations}
\label{sec:General}

In order to perform numerical simulations, we subdivide our system into a finite number of spherically symmetric shells. Within each shell, density, opacity and temperature are assumed to be constant. The contribution of shell $j$ to the radiation term at radius r is determined by:

\begin{equation}
A_{\mathrm r \leftarrow \mathrm j}=\int_{r_{\mathrm {jL}}}^{r_{\mathrm {jR}}}{r'}^2 \int_{0}^{\pi} \frac{e^{-\tau}}{D^2} \sin\left(\theta'\right) d\theta' dr'.
\label{eq:Aij}
\end{equation}

$r_{\mathrm {jL}}$ and $r_{\mathrm {jR}}$ are the inner and outer boundaries of shell j, respectively. In order to obtain the exchange coefficient between shells $i$ and $j$, the above expression would have to be integrated over shell $i$, i.e.
\begin{equation}
A_{\mathrm i \leftarrow \mathrm j}=\frac {\int_{m_{\mathrm {iL}}}^{m_{\mathrm {iR}}} A_{\mathrm r \leftarrow \mathrm j} \left(m\right) dm} {M_{\mathrm i}}.
\end{equation}
$m$ is the mass coordinate, $M_{\mathrm i}$ the mass contained within shell $i$. 

Unfortunately, the integration over shell $i$ can only be done numerically, so this method is slow. In addition, if we numerically integrate over $j$ as well (as in section \ref{sec:Num}), there are large inaccuracies for neighboring shells near the singularity that occurs at $r=r'$ for $\theta'=0$. For these reasons, we do not integrate over shell $i$, but instead we set $A_{\mathrm i \leftarrow \mathrm j}=A_{\mathrm r_{\mathrm i} \leftarrow \mathrm j}$, where $r_{\mathrm i}$ is the radius of shell $i$'s center.
This definition allows us to accurately calculate the radiation term at radius $r_{\mathrm i}$, but note that energy conservation between shells requires that $R_{\mathrm i \leftarrow \mathrm j} = -R_{\mathrm j \leftarrow \mathrm i}$. Since the radiation term is power per mass, this implies that $A_{\mathrm i \leftarrow \mathrm j} M_{\mathrm i} = A_{\mathrm j \leftarrow \mathrm i} M_{\mathrm j}$. With this definition, this is generally not fulfilled since the radiation term at the shell center is not equal to the shell average. In other words, when updating shell $i$, we calculate the effect of \textit{shell} $j$ on \textit{radius} $r_{\mathrm i}$, but when updating shell $j$, we calculate the effect of \textit{shell} $i$ on \textit{radius} $r_{\mathrm j}$. This is not symmetric between $i$ and $j$, which ultimately causes a violation of energy conservation. 

Alternatively, it is possible to use the following definition:

\begin{equation}
A_{\mathrm i \leftarrow \mathrm j}=\frac{A_{\mathrm r_{\mathrm i} \leftarrow \mathrm j} M_{\mathrm i} + A_{\mathrm r_{\mathrm j} \leftarrow \mathrm i} M_{\mathrm j}}{2 M_{\mathrm i}}
\end{equation}
The radiation term is now defined as an average between the two directions of transfer. This means the radiation term at $r_{\mathrm i}$ is no longer calculated as accurately, but this expression is symmetric to exchanging $i$ and $j$ and therefore conserves energy. 

The method we develop in the following sections is able to calculate the radiation term at a specific radius exactly in the limit of infinite resolution. Therefore, for the test calculations in this paper and to investigate the effect of different resolutions, we use the non-symmetrized definition and from now on we will refer to $A_{\mathrm r_{\mathrm i} \leftarrow \mathrm j}$ simply as $A_{\mathrm i \leftarrow \mathrm j}$. In a real application, the symmetrized definition may be preferred since it guarantees energy conservation.

\subsubsection{Constant-$\chi$-method}
\label{sec:ConstChi}

The fastest way of calculating $A_{\mathrm i \leftarrow \mathrm j}$ is to substitute $\tau=\overline \chi_{\mathrm {ij}} D$. This is an approximation: It assumes that $\overline \chi_{\mathrm {ij}}$, the proportionality constant between distance and optical depth, is a constant independent of $\theta'$. But unless $\chi$ is actually constant throughout the whole system, this is clearly not true. Consider again Figure \ref{fig:SpherSym}: If $\chi$ increases towards the center, as is usually the case, then rays going through the inner region will have larger values of $\chi_{\mathrm {ij}}$ than those who move only outwards.

It is ultimately up to the user which value to choose for $\overline \chi_{\mathrm {ij}}$; the simplest method, which we adopt here, is to take the value for $\theta'=0$. If $\chi$ increases towards the center, we expect this to cause an overestimation of $A_{\mathrm i \leftarrow \mathrm j}$ especially for the case $i<j$, and a somewhat smaller overestimation for $i>j$. Despite this inaccuracy, this method can be useful, because it allows us to solve the angle integral analytically and is therefore very fast. We start by expressing the distance D as a function of $r_{\mathrm i}$, $r_{\mathrm j}$ and $\theta'$:

\begin{align}
\begin{split}
&D^2={r_{\mathrm i}}^2+{r_{\mathrm j}}^2-2r_{\mathrm i} r_{\mathrm j} \cos\left(\theta'\right), \\
\Rightarrow &2D \frac{dD}{d\theta'} = 2 r_{\mathrm i} r_{\mathrm j} \sin\left(\theta'\right), \\
\Rightarrow &\sin\left(\theta'\right) d\theta'=\frac{D}{r_{\mathrm i} r_{\mathrm j}} dD.
\label{eq:D}
\end{split}
\end{align}

We insert this into the definition of $A_{\mathrm i \leftarrow \mathrm j}$ (equation \ref{eq:Aij}), but ignore the integration over shell $j$ for the moment and set $r'=r_{\mathrm j}$. Using $\tau=\overline \chi_{\mathrm {ij}} D$, we obtain

\begin{align}
\begin{split}
A_{\mathrm i \leftarrow \mathrm j}&=\frac{r_{\mathrm j}}{r_{\mathrm i}} \int_{D_{\mathrm {min}}}^{D_{\mathrm {max}}} \frac{e^{-\overline \chi_{\mathrm {ij}} D}}{D}  dD \Delta r_{\mathrm j}, \\
&=\frac{r_{\mathrm j}}{r_{\mathrm i}} \int_{\tau_{\mathrm {min}}}^{\tau_{\mathrm {max}}} \frac{e^{-\tau}}{\tau}  d\tau \Delta r_{\mathrm j}.
\label{eq:Dintegral}
\end{split}
\end{align}
The indices min and max refer to the minimum and maximum distances/optical depths between the radii $r_{\mathrm i}$ and $r_{\mathrm j}$ (see Figure \ref{fig:SpherSym}). The integral over $\tau$ can be reformulated using exponential integrals. The exponential integral of order n is defined as

\begin{equation}
E_n\left(x\right)=\int_x^\infty \frac{e^-t}{t^n} dt.
\end{equation}

Using this, equation \ref{eq:Dintegral} becomes

\begin{equation}
A_{\mathrm i \leftarrow \mathrm j}=\frac{r_{\mathrm j}}{r_{\mathrm i}} \left(E_{\mathrm 1}\left(\tau_{\mathrm {min}}\right) - E_{\mathrm 1}\left(\tau_{\mathrm {max}}\right)\right) \Delta r_{\mathrm j}.
\label{eq:EI}
\end{equation}

We have avoided the numerical integration over $\theta'$, and instead have to solve two exponential integrals. This is much faster since analytic formulas are available for the latter. The $\tau_{\mathrm {max}}$-term is negligible in optically thick systems, since in that case $E_{\mathrm 1}\left(\tau_{\mathrm {max}}\right) << E_{\mathrm 1}\left(\tau_{\mathrm {min}}\right)$.

We now consider the integration over shell $j$, so we replace the constant $r_{\mathrm j}$ with the variable $r'$. Based on equation (\ref{eq:EI}), the dependency on $r'$ is of the form $r' E_{\mathrm 1}\left(\tau\left(r'\right)\right)$. This can be integrated analytically using the relation between an exponential integral and its derivative, $\int E_{\mathrm n}\left(x\right) dx = -E_{\mathrm {n+1}}\left(x\right) + C$.

We express $r'$ as a function of $\tau$, assume $\frac{dr'} {d\tau}=$const., and perform integration by parts:

\begin{align}
\begin{split}
&\int r'\left(\tau\right) E_{\mathrm 1}\left(\tau\right) \frac{dr'} {d\tau} d\tau \\
=&\frac{dr'} {d\tau} \left(-r' E_{\mathrm 2} + \frac{dr'}{d\tau} \int  E_{\mathrm 2} d\tau\right) + C \\
=&\frac{dr'} {d\tau} \left(-r' E_{\mathrm 2} - \frac{dr'}{d\tau}  E_{\mathrm 3}\right) + C.
\label{eq:EIint}
\end{split}
\end{align}
The function $r'\left(\tau\right)$ differs depending on whether we are looking at $\tau_{\mathrm {min}}$ or $\tau_{\mathrm {max}}$, and whether shell $i$ is inside or outside of shell $j$:

\begin{equation}
r'=r_{\mathrm i} \pm \frac{\tau_{\mathrm {min}}}{\overline \chi_{\mathrm {ij}}} = r_{\mathrm i} + \frac{\tau_{\mathrm {max}} - 2 \overline \chi_{\mathrm {ij}} r_{\mathrm i}}{\overline \chi_{\mathrm {ij}}}.
\label{eq:rvontau}
\end{equation}
Here, $\pm$ means + in the case of $r_{\mathrm i}<r'$ and - in the case of $r_{\mathrm i}>r'$.
In every case, $\frac{dr'}{d\tau}$ is either $\overline \chi_{\mathrm {ij}}^{-1}$ or $-\overline \chi_{\mathrm {ij}}^{-1}$, validating the above assumption $\frac{dr'}{d\tau}=$const.

To summarize, we have 3 different relations between $r'$ and $\tau$: $r'_-\left(\tau_{\mathrm {min}}\right)$, $r'_+\left(\tau_{\mathrm {min}}\right)$ and $r'\left(\tau_{\mathrm {max}}\right)$.

This situation arises because the minimum optical depth (case $\theta'=0$) either increases or decreases with $r'$ depending on whether $r'>r_{\mathrm i}$ or $r'<r_{\mathrm i}$, but the maximum optical depth (case $\theta'=\pi$) always increases with $r'$. 

\subsubsection{Numerical Angle Integration}
\label{sec:Num}
We now abandon the assumption of a constant $\overline \chi_{\mathrm {ij}}$, which makes it impossible to solve the angle integral analytically. For numerical integration, one may use either variant (\ref{eq:BasicEqOrig}) or (\ref{eq:BasicEq}) of the basic equation. We choose the latter for consistency with the previous section, but the method discussed here is essentially independent of the formulation used.

Recall that our definition of $A_{\mathrm i \leftarrow \mathrm j}$ is

\begin{equation}
A_{\mathrm i \leftarrow \mathrm j}=\int_{r_{\mathrm {jL}}}^{r_{\mathrm {jR}}}{r'}^2 \int_{0}^{\pi} \frac{e^{-\tau}}{D^2} \sin\left(\theta'\right) d\theta' dr'.
\label{eq:angleint}
\end{equation}

We consider rays emitted from point $\vec r$ in directions between $\tilde \theta=0$ and $\pi$ (see again Figure \ref{fig:SpherSym} for a visualization of the situation). It is important to note that, unless $r=0$, the direction angle $\tilde \theta$ is different from the polar angle $\theta'$. $\tau$ depends on $D$ and $\tilde \theta$, so if we have a total of $n$ shells and the number of direction angles considered is $a$, for each individual frequency we create 2-dimensional arrays of size $n*a$ in which to store the values of $\tau$, $D$ and $\theta'$ for each combination of $j$ and $\tilde \theta$. Once that is done, the integral can be calculated numerically. In addition, implementing the numerical integration over the radius of shell $j$ is straightforward: Instead of stopping only at each shell center and writing $\tau$, $D$ and $\theta'$ there, we do so multiple times within each shell. 

The rest of this section describes how to obtain these arrays. For each direction angle $\tilde \theta$, we start from position $\vec r$ and then move along the ray from shell to shell until we reach the outer boundary. We will now summarize the algorithm for an individual direction: 

\begin{figure}
		\plotone{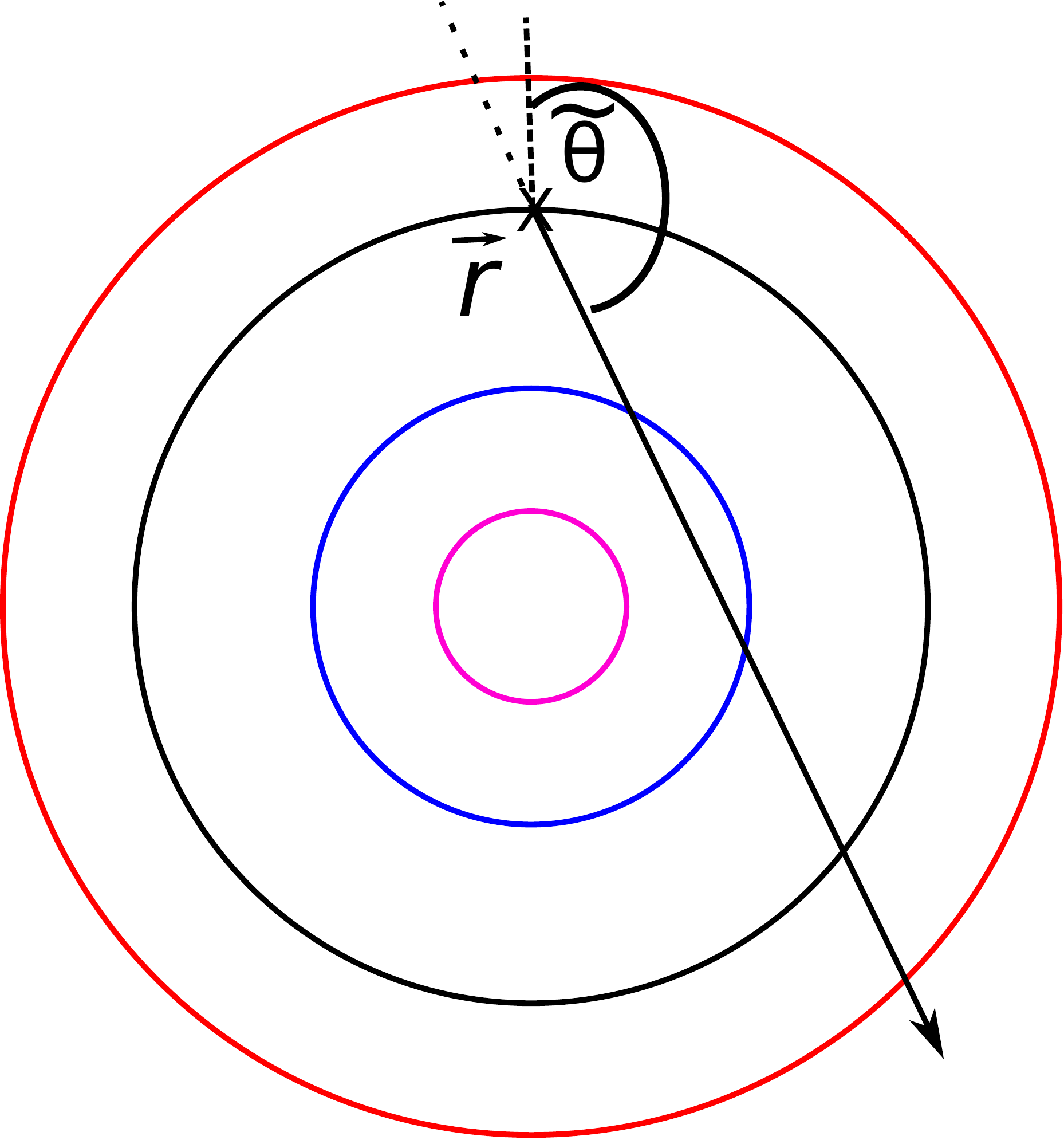}
	\caption{We see here a ray moving inward from position $\vec r$. The 4 different cases in section \ref{sec:Num} are illustrated, with the colored shells representing different target radii: \newline
	\textcolor[rgb]{1,0,0}{Case 1:} Two solutions: The ray crosses the shell once in the positive $x$-direction, and once more in the negative direction which we ignore. \newline
	\textcolor[rgb]{0,0,1}{Case 2:} Two positive solutions: The ray crosses the shell once moving inwards, and once again coming out, both in the positive $x$-direction. \newline
	\textcolor[rgb]{1,0,0.8}{Case 3:} No solution: The shell is too small for this ray to hit it at all. \newline
	Case 4: Current and target radius are equal so we have a trivial solution ($dx=0$), and a second solution by moving through the inner region. \newline
	}
	\label{fig:Num}
\end{figure}

First, we define variables representing the current position (initially $\vec r$), distance and optical depth (both initially 0). Then, we move along the ray from one shell to the next, all the while updating distance and optical depth and writing the results in our arrays. $\theta'$ can easily be calculated from the position. This continues until reaching the outer boundary. This scheme is similar to the ``impact parameter'' method introduced by \citet{Hummer1971}, in the sense that we move from shell to shell along a ray. In that method, parallel rays at different distances (impact parameters) from the center are considered. The calculation then jumps from one intersection of such a ray with a discrete spherical shell to the next in order to track the changing intensity. This can then be used to calculate the moments of radiation and the Eddington factor. The main difference in our method is the fact that we do not follow the intensity, but instead we calculate the distance and optical depth between any two shells in a certain direction, in order to obtain our coefficients for the radiative exchange between those two.

To move along the ray, at every step we must define a ``target radius'' to reach with the next step. For outward-moving rays ($\tilde \theta < \frac{\pi}{2}$), the radial coordinate is always increasing, so the target radius is always larger than the current radius. For inward-moving rays, such as the one depicted in Figure \ref{fig:Num}, the situation is more complicated since the radius decreases down to a minimum before increasing again. In addition, the ray crosses smaller shells not once, but either twice or never (the blue and green shells in Figure \ref{fig:Num}).

If our current position is $\left(x_{\mathrm c},y_{\mathrm c}\right)$ at radius $r_{\mathrm c}$, we wish to move along the direction of the ray by a distance $\Delta D$ until we reach the target radius $r_{\mathrm t}$: 

\begin{align}
&\Delta y=\frac{\Delta x}{tan\left(\tilde \theta\right)},
&\Rightarrow r_{\mathrm t}^2=\left(x_{\mathrm c}+\Delta x\right)^2 + \left(y_{\mathrm c}+\frac{\Delta x}{tan\left(\tilde \theta\right)}\right)^2. 
\end{align}

Defining $g=1 / tan\left(\tilde \theta\right)$, after some manipulation we obtain the following quadratic equation for $\Delta x$:

\begin{align}
&\Delta x^2 + \frac{2x_{\mathrm c}+2y_{\mathrm c} g}{1+g^2} \Delta x + \frac{r_{\mathrm c}^2-r_{\mathrm t}^2}{1+g^2} = 0, \\
&\Rightarrow \Delta x_{\mathrm {1,2}}=-\frac{x_{\mathrm c}+y_{\mathrm c} g}{1+g^2} \pm \sqrt{\left(\frac{x_{\mathrm c}+y_{\mathrm c} g}{1+g^2}\right)^2 - \frac{r_{\mathrm c}^2-r_{\mathrm t}^2}{1+g^2}}.
\end{align}

This equation tells us how far in $x$-direction we need to move along the ray to reach the target radius. There are 4 possible cases, whose geometric meaning is depicted in Figure \ref{fig:Num}:

\begin{enumerate}
\item $r_{\mathrm t} > r_{\mathrm c}$: There are 2 solutions, one positive and one negative. Since $0 \leq \tilde \theta \leq \pi$, all the rays we consider move in the positive $x$-direction, so we choose the positive solution.
\item $r_{\mathrm t} < r_{\mathrm c}$ and the value under the root is positive: There are 2 positive solutions. We choose the smaller of the two since otherwise we would go through the whole inner region in a single step.
\item $r_t < r_c$ and the value under the root is negative: There is no real solution. This means that the target radius is not reached by this ray. We set $r_{\mathrm t}=r_{\mathrm c}$ and recalculate.
\item $r_{\mathrm t}=r_{\mathrm c}$: One solution is 0 and the other one is $-\left(2x_{\mathrm c}+2y_{\mathrm c} g\right) / \left(1+g^2\right)$. We choose the latter.
\end{enumerate}

Note that for outward-moving rays ($\tilde \theta < \pi / 2$), only case 1 occurs since the radius is always increasing.

\subsubsection{The Boundary Term}
\label{sec:BT}
We have not yet considered the second term in the basic equation (\ref{eq:BasicEq}). This 'boundary term' is given by

\begin{align}
\begin{split}
&\kappa \int_\Omega \left(S_{\mathrm O}-S\right) e^{-\tau_{\mathrm B}} d\Omega \\
=&2 \pi \kappa \int_{\tilde \theta}\left(S_{\mathrm O}-S\right) e^{-\tau_{\mathrm B}} \sin\left(\tilde \theta\right) d\tilde \theta. 
\end{split}
\end{align}

The index O here means the value at the outer boundary, while the index B means ``from radius r to the outer boundary''. We shall not consider the possibility of a $\phi$-dependent $S_{\mathrm O}$ here, so that we can eliminate the azimuthal angle and treat the system as 2-D. Including the boundary term in the numerical angle integration (\ref{sec:Num}) is straightforward: We simply add an additional element to our array for $\tau$ at the radius of the outer boundary and integrate numerically.

For the constant-$\chi$-method, we must numerically integrate as well since, even if $S_{\mathrm O}$ is isotropic, there is no analytic solution. $\tau_{\mathrm B}$ can be expressed as $\overline \chi_{\mathrm {B}} D_{\mathrm B}$, and 
\begin{equation}
D_{\mathrm B}=\sqrt{r^2+r_{\mathrm O}^2-2rr_{\mathrm O} \cos\left(\theta'\right)}.
\end{equation}

Here, the polar angle $\theta'$ appears again so we require a relation between $\theta'$ and $\tilde \theta$.

$\tilde \theta$ is simply the polar angle of a coordinate system centered at $\vec r$, and $D$ is equivalent to the radial coordinate in that system. Since $\vec r$ lies on the $y$-axis, for any point ($x'$,$y'$):

\begin{align}
&x' = D \sin \left(\tilde \theta\right), \\
&y' = D \cos \left(\tilde \theta\right) + r, \\
&\Rightarrow \tilde \theta=\mathrm {atan} \left(\frac{x'}{y'-r}\right).
\end{align}

Numerical integration is performed over $\theta'$. For each $\theta'$, calculate $D_{\mathrm B}$, $x'$ and $y'$, which leads to $\tilde \theta$ and ultimately $\Delta \tilde \theta \left(\theta'\right)$, allowing numerical integration.

While the lack of an analytic solution makes this comparatively slow, it hardly affects the total calculation time because the boundary term only needs to be calculated once per shell, not $n$ times per shell.

Nevertheless, it is possible to find an analytic solution by approximating $\tilde \theta = \theta'$. This approximation only holds if $r<<r_{\mathrm O}$, but this can be reasonable if, for instance, one is interested in the behavior of a small object inside a large, optically thin envelope.

If $S_{\mathrm O}$ does not depend on $\tilde \theta$, we can use equation (\ref{eq:D}) and $\tau=\overline \chi_{\mathrm {B}} D_{\mathrm B}$ to write

\begin{align}
\begin{split}
&\int_{\tilde \theta} e^{-\tau_{\mathrm B}} \sin\left(\theta'\right) d\theta' = \frac{1}{rr_{\mathrm B} \overline \chi_{\mathrm {B}}^2} \int_{\tau_{\mathrm Bmin}}^{\tau_{\mathrm {Bmax}}} e^{-\tau_{\mathrm B}} d\tau_{\mathrm B} \\
=&\frac{1}{rr_{\mathrm O}\overline \chi_{\mathrm {B}}^2}\left(\left(1+\tau_{\mathrm {Bmin}}\right)e^{-\tau_{\mathrm {Bmin}}} - \left(1+\tau_{\mathrm {Bmax}}\right) e^{-\tau_{\mathrm {Bmax}}}\right).
\end{split}
\end{align}

\subsection{Optically Thick Shells}
\label{sec:thick}

So far, we have implicitly assumed that the individual shells in our system are sufficiently optically thin that the radiative exchange between different points of the same shell is negligible. The method outlined in the previous sections no longer works if that assumption does not hold. Most of the radiative exchange then should happen within a shell, but in our method, the temperature within each shell is constant, so this internal exchange is by definition zero. This section explains how to expand the method to overcome this problem.

We make use of the constant-$\chi$-approximation and use the expression from equation (\ref{eq:EI}) for $A_{\mathrm i \leftarrow \mathrm j}$. We modify the definition to include $\left(S'-S\right)$, which is no longer zero throughout the shell:

\begin{equation}
A_{\mathrm i \leftarrow \mathrm i}=\frac{1}{r} \int_{r_{\mathrm {iL}}}^{r_{\mathrm {iR}}} r' \left(S'-S\right) \left(E_{\mathrm 1}\left(\tau_{\mathrm {min} }\right) - E_{\mathrm 1}\left(\tau_{\mathrm {max} }\right)\right) dr'.
\end{equation}

Concerning the applicability of the constant-$\chi$-approximation: If a single shell is optically intermediate or thick already, then the effect of rays which have to move through the interior region will generally be negligible. Because of this, assuming that the direction-independent $\overline \chi_{\mathrm {ii}}$ is equal to  $\chi_{\mathrm i}$ is generally a good approximation. However, we recognize that this can introduce some inaccuracy in certain atypical scenarios, such as an optically intermediate shell surrounding an optically thin interior region.

We abbreviate $\left(S'-S\right)$ as a function $f\left(r'\right)$ and interpolate it within shell $i$. We consider here a second-order polynomial of the form

\begin{equation}
f\left(r'\right)=a\left(r'-r_{\mathrm i}\right)^2 + b\left(r'-r_{\mathrm i}\right) + c.
\end{equation}

Since $f\left(r_{\mathrm i}\right)=0$, it follows that $c=0$, and the other two coefficients can be calculated from the known values $f\left(r_{\mathrm {i+1}}\right)$ and $f\left(r_{\mathrm {i-1}}\right)$.

To integrate, $r'$ and $f$ must be expressed as functions of $\tau_{\mathrm {min} } / \tau_{\mathrm {max} }$. We have already done this for $r'$ in equation (\ref{eq:rvontau}). Similarly for $f$, we obtain

\begin{align}
&f_\pm\left(\tau_{\mathrm {min} }\right)=\frac{a}{\chi_{\mathrm i}^2} \tau_{\mathrm {min} }^2 \pm \frac{b}{\chi_{\mathrm i}} \tau_{\mathrm {min} }, \\
\begin{split}
&f\left(\tau_{\mathrm {max} }\right)=\frac{a}{\chi_{\mathrm i}^2} \left(\tau_{\mathrm {max} }-2 \tau_{\mathrm C} \right)^2 + \frac{b}{\chi_{\mathrm i}} \left(\tau_{\mathrm {max} }-2 \tau_{\mathrm C} \right) \\
=&\frac{a}{\chi_{\mathrm i}^2} \tau_{\mathrm {max} }^2 + \left(\frac{b}{\chi_{\mathrm i}} - \frac{4a \tau_{\mathrm C}}{\chi_{\mathrm i}^2}\right) \tau_{\mathrm {max} } + \left(\frac{4a \tau_{\mathrm C}}{\chi_{\mathrm i}^2}-\frac{2b}{\chi_{\mathrm i}}\right) \tau_{\mathrm C}.
\end{split}
\end{align}
Again, $\pm$ means + for $r'>r_{\mathrm i}$ and - for $r'<r_{\mathrm i}$, and $\tau_{\mathrm C}=\chi_{\mathrm i} r_{\mathrm i}$ is the optical depth from the center of shell $i$ to the center of the system if $\chi$ were constant.

We split the integral into parts:

\begin{align}
\begin{split}
&\int_{r_{\mathrm {iL}}}^{r_{\mathrm {iR}}} r' f\left(r'\right) \left(E_{\mathrm 1}\left(\tau_{\mathrm {min} }\right) - E_{\mathrm 1}\left(\tau_{\mathrm {max} }\right)\right) dr' \\
=&+\frac{dr'_-}{d\tau_{\mathrm {min} }} \int_{\tau_{\mathrm {minL}}}^0 r'_-\left(\tau_{\mathrm {min} }\right) f_-\left(\tau_{\mathrm {min} }\right) E_{\mathrm 1}\left(\tau_{\mathrm {min} }\right) d\tau_{\mathrm {min} } \\
&+\frac{dr'_+}{d\tau_{\mathrm {min} }} \int_0^{\tau_{\mathrm {minR}}} r'_+\left(\tau_{\mathrm {min} }\right) f_+\left(\tau_{\mathrm {min} }\right) E_{\mathrm 1}\left(\tau_{\mathrm {min} }\right) d\tau_{\mathrm {min} }\\
&-\frac{dr'}{d\tau_{\mathrm {max} }} \int_{\tau_{\mathrm {maxL}}}^{\tau_{\mathrm {maxR}}} r'\left(\tau_{\mathrm {max} }\right) f\left(\tau_{\mathrm {max} }\right) E_{\mathrm 1}\left(\tau_{\mathrm {max} }\right) d\tau_{\mathrm {max} } \\
\end{split}
\end{align}

The leading factors are $\chi_{\mathrm i}^{-1}$ or $-\chi_{\mathrm i}^{-1}$. Since the shell-internal exchange only matters for optically thick systems, the third term is usually negligible unless the shell is optically intermediate and we are so close to the center that $E_{\mathrm 1}\left(\tau_{\mathrm {max} }\right)$ is comparable to $E_{\mathrm 1}\left(\tau_{\mathrm {min} }\right)$.

After inserting the expressions for the various functions $r'$ and $f'$, we see that these integrals all have the following form:

\begin{align}
\begin{split}
&\int \left(\overline a \tau^2 + \overline b \tau + \overline c\right)\left(\overline d \tau + \overline e\right) E_{\mathrm 1}\left(\tau\right) d\tau  \\
&=\overline a \overline d I_{\mathrm {3,1}} + \left(\overline a \overline e + \overline b \overline d\right) I_{\mathrm {2,1}} + \left(\overline b \overline e + \overline c \overline d\right) I_{\mathrm {1,1}} + \overline c \overline e I_{\mathrm {0,1}} d\tau.
\end{split}
\end{align}

The coefficients $\overline a$ through $\overline e$ depend on the integral in question and 

\begin{align}
\begin{split}
I_{\mathrm {n,m}}\left(\tau\right)&=\int \tau^n E_{\mathrm m} d\tau \\
&=-\tau^n E_{\mathrm {m+1}} + n \int \tau^{n-1} E_{\mathrm {m+1}} d\tau  \\
&=-\tau^n E_{\mathrm {m+1}} + n I_{\mathrm {n-1,m+1}} 
\end{split}
\end{align}

This allows us to solve all of these integrals recursively, with $\int I_{\mathrm {0,m}}\left(\tau\right)=-E_{\mathrm {m+1}} \left(\tau\right) + C$.

\subsection{Method Summary}
In our method, there are three terms contributing to the total radiation term:

\begin{enumerate}
\item The radiative exchange between different shells (``External Term'')
\item The radiative exchange within the same shell (``Internal Term'')
\item The radiative exchange with the background radiation field (``Boundary Term'')
\end{enumerate}

The optical depth structure and the resolution determine which of these terms dominate and which are negligible. In a completely optically thin system, the boundary term dominates everywhere. When the system becomes optically thick, but individual shells are still optically thin, the external term dominates the optically thick region, while the boundary term is still important in the outer regions. Finally, if the optical depth becomes so large that individual shells become optically thick, the internal term dominates in that region.

In terms of computational effort, the external term is by far the slowest since even with the fast method (constant-$\chi$-approximation, section \ref{sec:ConstChi}), it is $O\left(n^2\right)$ for each frequency, while the other two are $O\left(n\right)$. With numerical integration over a angles as in section \ref{sec:Num}, it becomes $O\left(an^2\right)$, and if splitting shell $j$ into $x$ subshells for numerical integration, it increases to $O\left(axn^2\right)$.

The internal term could in theory be made obsolete by choosing a sufficiently high resolution that all shells remain optically thin. But in realistic simulations of astrophysical objects such as stars, the opacity and density in the inner region are so high that this is impossible in practice. In addition, it is not desirable since the internal term is much faster than the external term.

We summarize the main advantages of our method:

\begin{enumerate}
\item It is accurate for both optically thin and thick systems and smoothly transitions between the two.
\item It is fast, especially when using the constant-$\chi$-approximation. Significant increases in speed can be achieved by neglecting terms (especially the external term) in regions where they are not relevant.
\item The constant-$\chi$-method can be combined with numerical angle integration to achieve the best compromise between speed and accuracy (see section \ref{sec:Combo} for details on this).
\item If we ignore the weak temperature-dependency of $\kappa_{\nu}$, then the $A_{\mathrm i \leftarrow \mathrm j}$ for the external and boundary term depend only on the density profile but not on the temperature. In systems where $\rho\left(r\right)$ changes much more slowly than $T\left(r\right)$, these $A_{\mathrm i \leftarrow \mathrm j}$ only need to be recalculated very rarely, making the method extremely fast.
\end{enumerate}

\section{Test Calculations}
\label{sec:TestCalcs}

\subsection{Homogeneous Extinction Coefficient}
\label{sec:HomChi}
\subsubsection{Constant Temperature}
\label{sec:ConstT}
We begin with the simplest possible case: A static, fully homogeneous medium with constant temperature. Of course, the radiation term is zero everywhere in this case, but nevertheless we can assess the accuracy of our method by looking at the radiative cooling only. Even in more complicated cases, the radiative cooling is known to be $4 \pi \kappa S$, giving us an analytic result to compare our method to. Unfortunately, an analytic solution for the heating and therefore the total radiation term is generally not available except in this trivial case and in the thermal relaxation test (section \ref{sec:TRM}).

The accuracy of our method depends on the optical depth structure of the system. In practice, this is different for every frequency, but for our test calculations, it is sufficient to look at a single optical depth array. We therefore employ the gray approximation, using the Planck opacity and expressing S as $\sigma T^4 / \pi$.

Instead of the radiation term, we calculate the total radiative cooling by simply setting $S'$ and $S_{\mathrm O}$ to zero in all the equations. The optical depth per shell and the number of shells determine which of the three terms in our method dominate. We consider optically thick systems with three different values of $\tau_{\mathrm S}$ (the optical depth per shell in radial direction):

\begin{itemize}
\item $\tau_{\mathrm S}=0.1$: The external term dominates most of the system, but the boundary term is important in the outer region.
\item $\tau_{\mathrm S}=1$: Internal and external terms are comparable. The boundary term is negligible except in the outermost shell.
\item $\tau_{\mathrm S}=10$: The internal term dominates everywhere.
\end{itemize}

\begin{figure}[htbp]
		\plotone{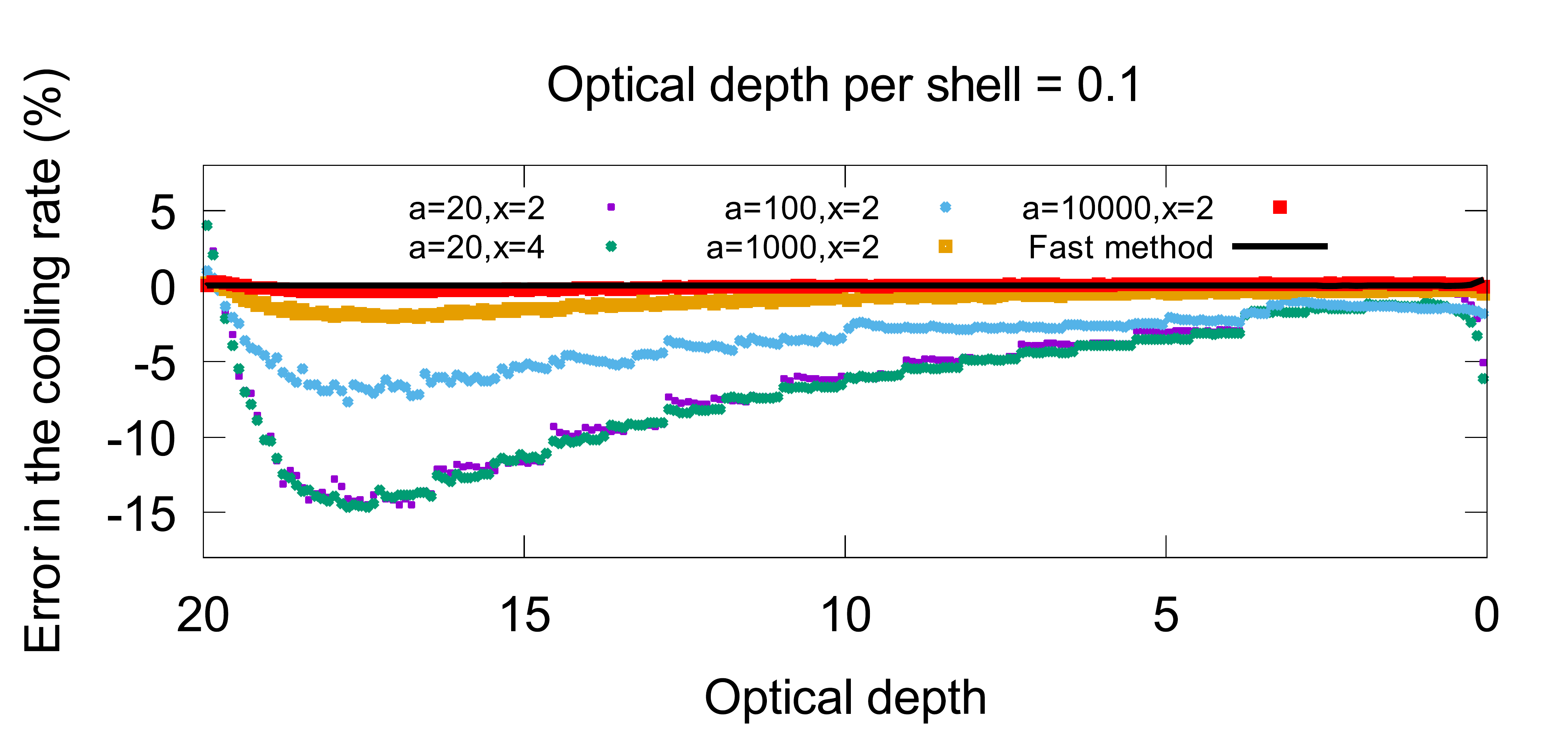}
		\plotone{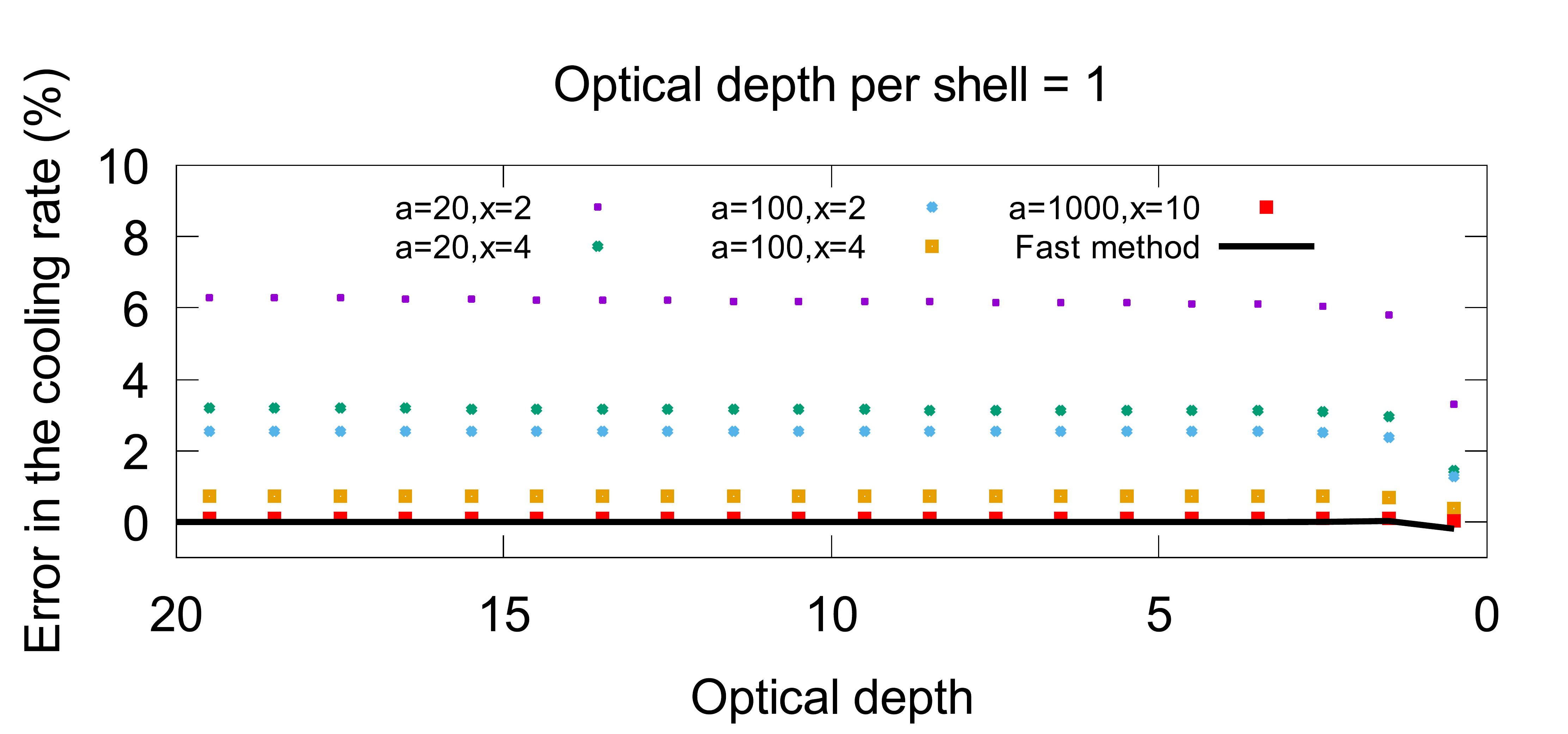}
		\plotone{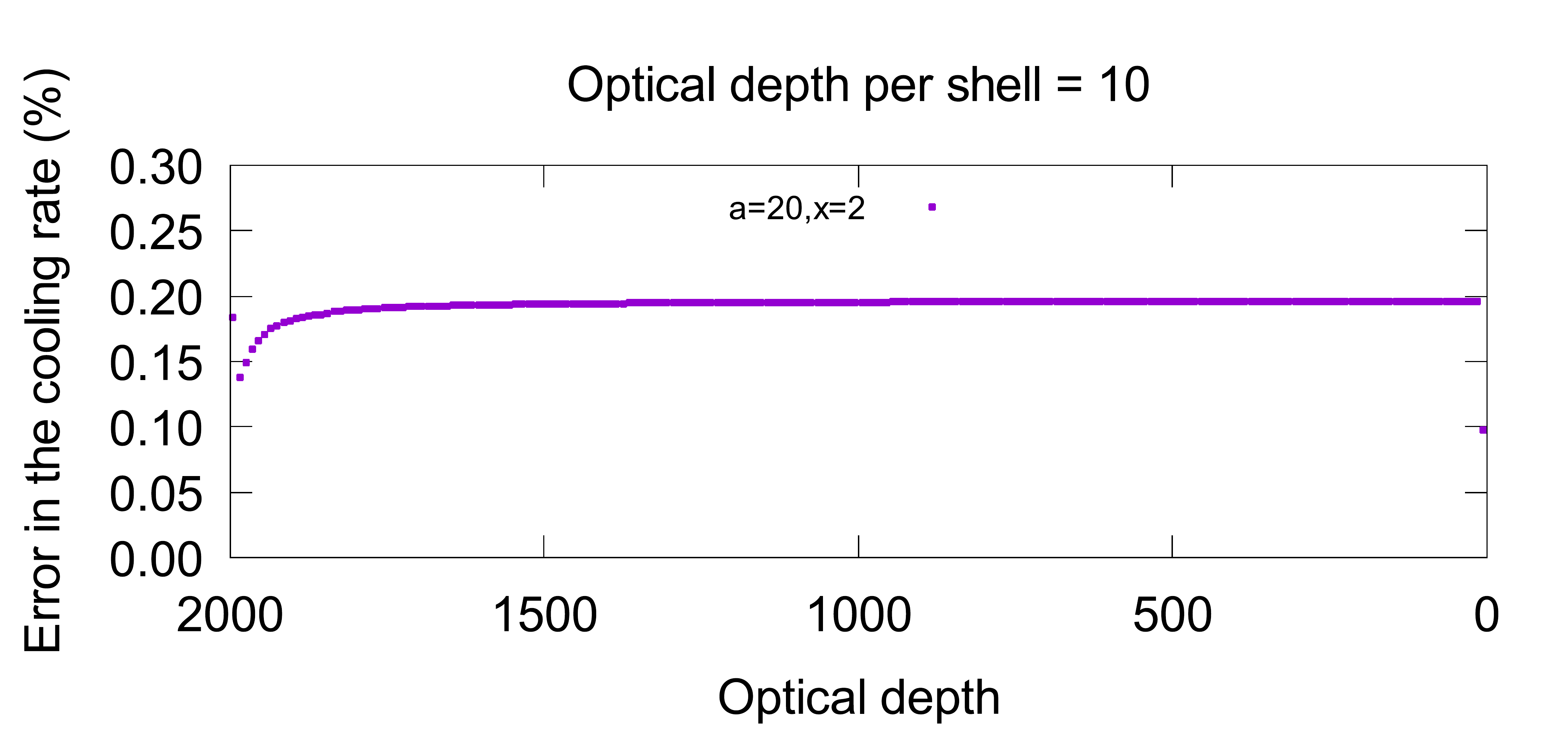}
	\caption{Homogeneous system: Relative error in the cooling rate for different optical depths per shell, and for various angular and sub-shell resolutions. The black line shows the result for the fast (constant-$\chi$) method. For the numerical integration cases, $a$ is the number of angles, $x$ is the number of points per shell.}
	\label{fig:ConstChiCooling}
\end{figure}

We discuss the relative error in the cooling rates (Figure \ref{fig:ConstChiCooling}): As expected, the constant-$\chi$ method is nearly exact in this case, since the system's density and opacity are homogeneous. The only inaccuracy comes from the limited angular resolution of the boundary term (recall from section \ref{sec:BT} that the boundary term is integrated numerically over all angles $\theta'$, even for the fast method). This resolution should be chosen for each shell individually, since for shells very close to the outer boundary, the dependency of distance and optical depth on $\theta'$ is much stronger than for shells further inside. We do not investigate the effect of this resolution in detail since its impact on the overall calculation time is very small.

Let us now turn to the numerical integration. We investigate the dependency of the error on two resolutions, namely the angular resolution (number of angles: $a$) and the radial resolution in shell $j$ (number of points considered in shell $j$: $x$).

Consider first the upper two plots. Both have a total optical depth of 20 and a radial size of 1 per shell, but the optical depth per shell and the number of shells differ.

For $\tau_{\mathrm S}=1$, in the lowest resolution case ($a=20$, $x=2$) the numerical integration overestimates the result by about 6\%. Both increasing $a$ and increasing $x$ helps to reduce this error. In contrast, for $\tau_{\mathrm S}=0.1$, the angular resolution alone determines the accuracy. This is expected since for optically thin shells, the exponential term $e^{-\tau}$ hardly changes between the inner and outer boundary of the shell, so once individual (sub)shells are optically thin, further increase of $x$ has no effect.

We further notice that in the case of optically thin shells, the error is not constant but increases towards the system center up to a maximum, and then decreases again. In addition, higher angular resolution is required to achieve a similar accuracy as the $\tau_{\mathrm S}=1$ case. To understand this behavior, refer to case 3 in Figure \ref{fig:Num}: When calculating the exchange between shell $i$ and a smaller shell $j$, only a subset of all direction angles $\tilde \theta$ actually interact with shell $j$. This subset becomes smaller as the ratio of $r_{\mathrm j}$ to $r_{\mathrm i}$ becomes smaller, effectively decreasing the angular resolution, until it eventually becomes zero. In other words, the exchange with shells smaller than $i$ becomes first inaccurate and then neglected, causing an underestimation of $A_{\mathrm i \leftarrow \mathrm j}$. This also explains why the error decreases again near the center, since shells there mainly interact with shells larger than themselves. In the case of $\tau_{\mathrm S}=1$, this error does not appear since only the first few neighboring shells contribute, and their radius is very similar to $r_{\mathrm i}$.

The magnitude of this error is limited by the fact that smaller shells also have a smaller surface area, decreasing their overall contribution, but care must be taken: Just because some shell $j$ is negligible for the cooling (i.e. it only absorbs a very small amount of the radiation emitted by shell $i$), does not necessarily mean that it is negligible for the heating as well, since the shell may be very hot. In such a case, it must be ensured that the angular resolution is high enough to capture all shells that are relevant for the heating.

Finally, the case of $\tau_{\mathrm S}=10$ shows only a tiny error even with very low angular and radial resolution. This is understandable since the internal term, which is the only term that matters in this case, depends on neither of these. It does not matter how accurately the external and boundary terms are calculated because they are negligible anyway.

\subsubsection{Thermal Relaxation Test}
\label{sec:TRM}

\begin{figure}[htbp]
		\plotone{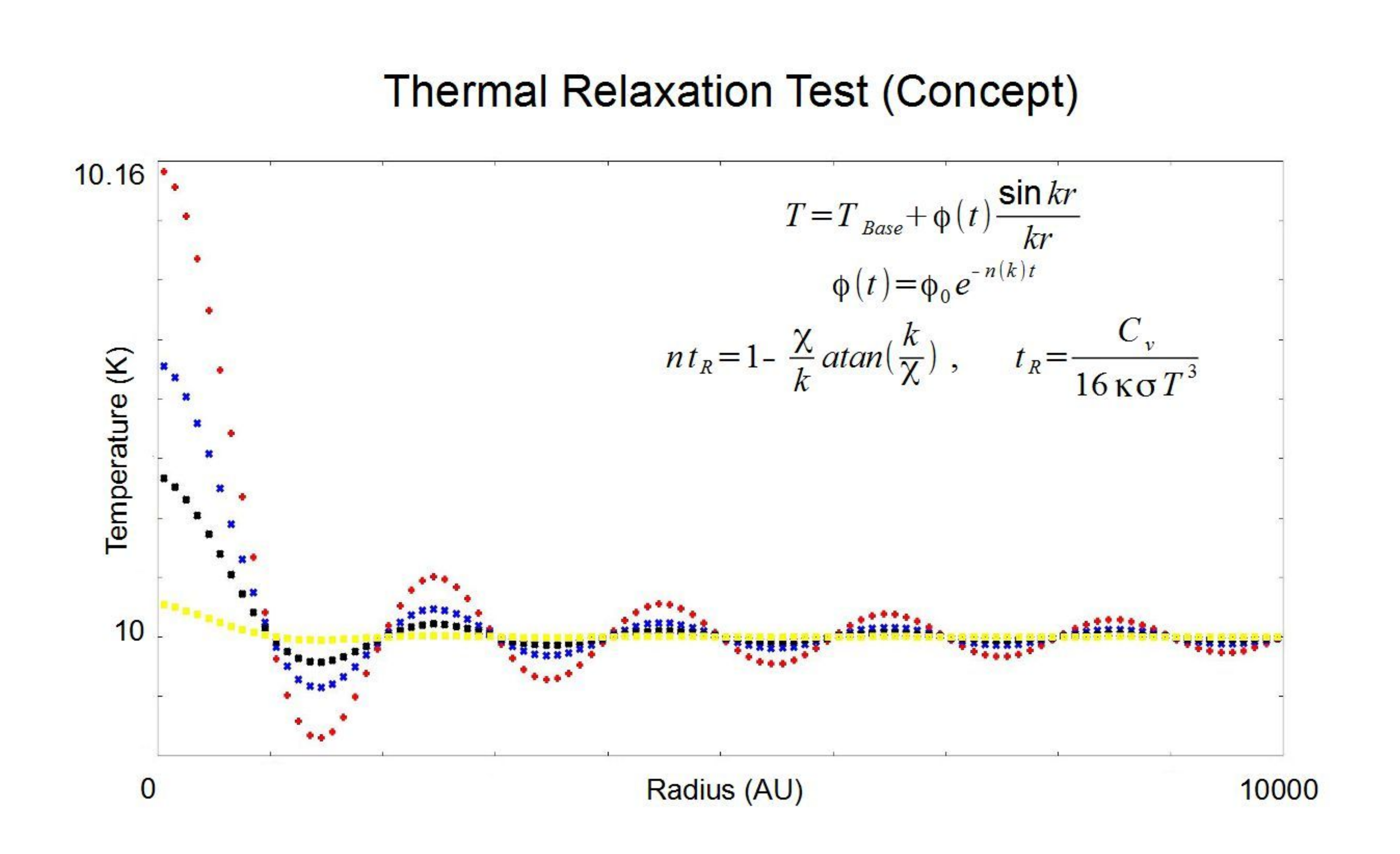}
	\caption{Concept of the thermal relaxation mode in spherical symmetry. An initial temperature perturbation with the shape of the zeroth-order spherical Bessel function decays over time (various time steps shown in different colors). The decay rate $n(k)$ can be calculated analytically. For a detailed derivation, see \citet{inu98}.}
	\label{fig:TRMConcept}
\end{figure}

A small temperature perturbation in an otherwise homogeneous system will be smoothed out over time by radiation (``thermal relaxation''), and in this special case an analytic solution can be found. This was first done by \citet{TRM} for a plane wave, and \citet{inu98} showed that the thermal relaxation rate for a spherical wave is exactly the same, provided that the initial perturbation has the form of the zeroth-order spherical Bessel function $j_{\mathrm 0}\left(r\right)=\sin \left(kr\right)/ \left(kr\right)$ (see Figure \ref{fig:TRMConcept} for the concept). This was also confirmed by \citet{Stamatellos2007}, in which they applied their radiative transfer scheme for SPH simulations to a spherical thermal relaxation test.

\begin{figure}[tbp]
	\plotone{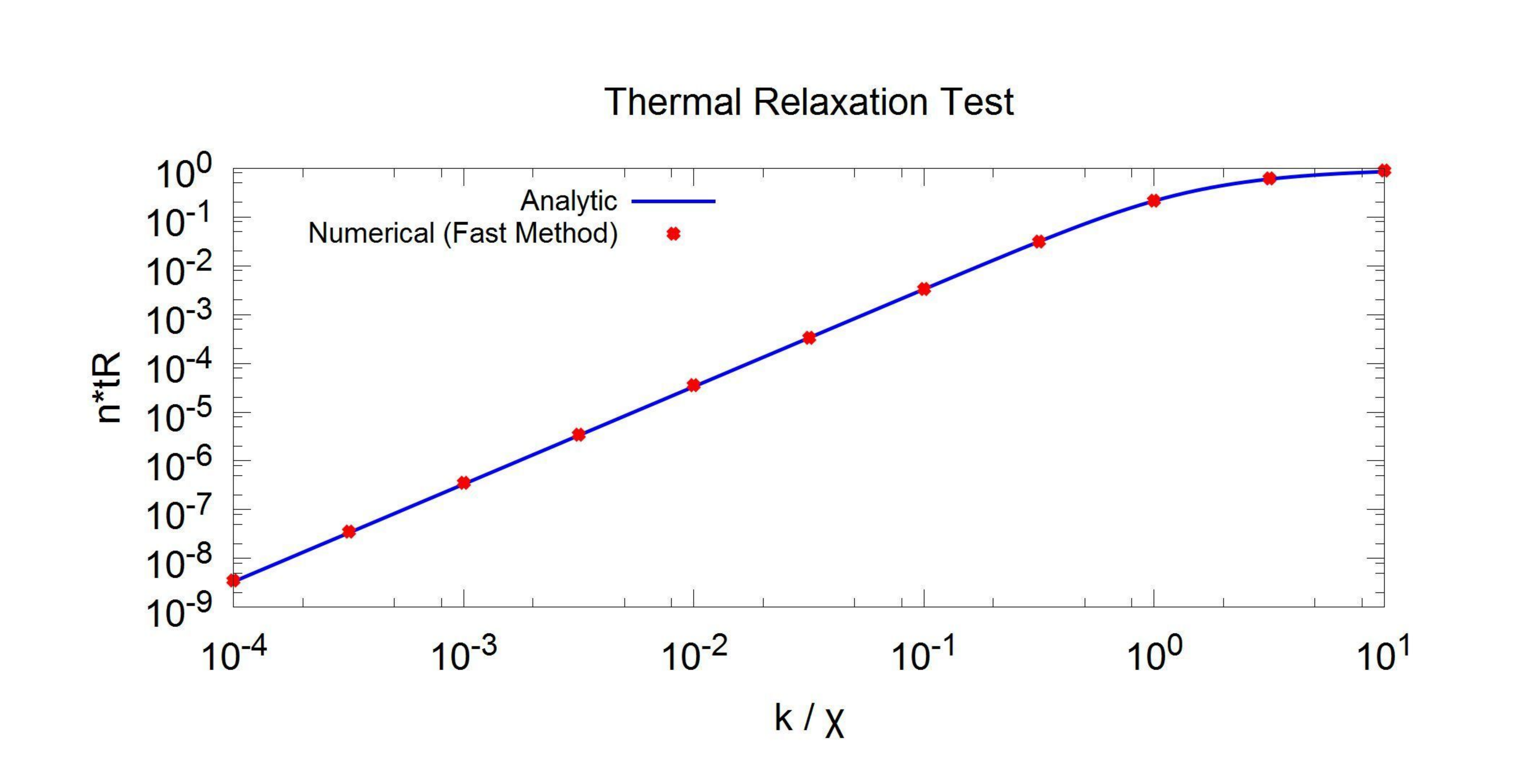}
	\caption{Result of the thermal relaxation test. This plot was obtained with the constant-$\chi$-method, but results for the numerical angle integration are indistinguishable even at low resolutions. These results include both calculations with optically thin and optically thick shells. For very large values of $k / \chi$, the result converges towards simple optically thin cooling.}
	\label{fig:TRM}
\end{figure}

Similarly, we test our method by superimposing small perturbations on a base temperature of 10K. We then measure the decay rate for various values of $k / \chi$. Figure \ref{fig:TRM} shows very good agreement with the analytic result, confirming that our method works in the case of a homogeneous system.

\subsection{Variable Extinction Coefficient: Protostellar Collapse Calculations}
\label{sec:VarChi}

As an example for a system where $\chi$ is not constant, we consider the gravitational collapse of a molecular cloud core. This is the mechanism by which high-density cores within molecular clouds form into stars. In a future paper, we will apply our method to detailed simulations of this process, but for now, we simply use it as a test for the method. The system transitions from an initially optically thin, constant-temperature, constant-$\chi$ state into a a system with a small, hot, optically thick core surrounded by a cold and thin envelope. Therefore, at different stages and different radii over the course of the collapse, all of the three terms in our method become important. For the hydrodynamics part of the simulation, we use an ideal gas equation of state with $\gamma=5/3$ and a second order Godunov scheme based on \citet{Colella1984}. We start from the following initial conditions:

\begin{itemize}
\item Core mass: $1M_{*}$
\item Initial radius: $10^4$ AU
\item Initial and background temperature: 10 K
\item Initial (constant) number density: $10^4 \mathrm{cm^{-3}}$
\end{itemize}

\begin{figure}[tbp]
		\plotone{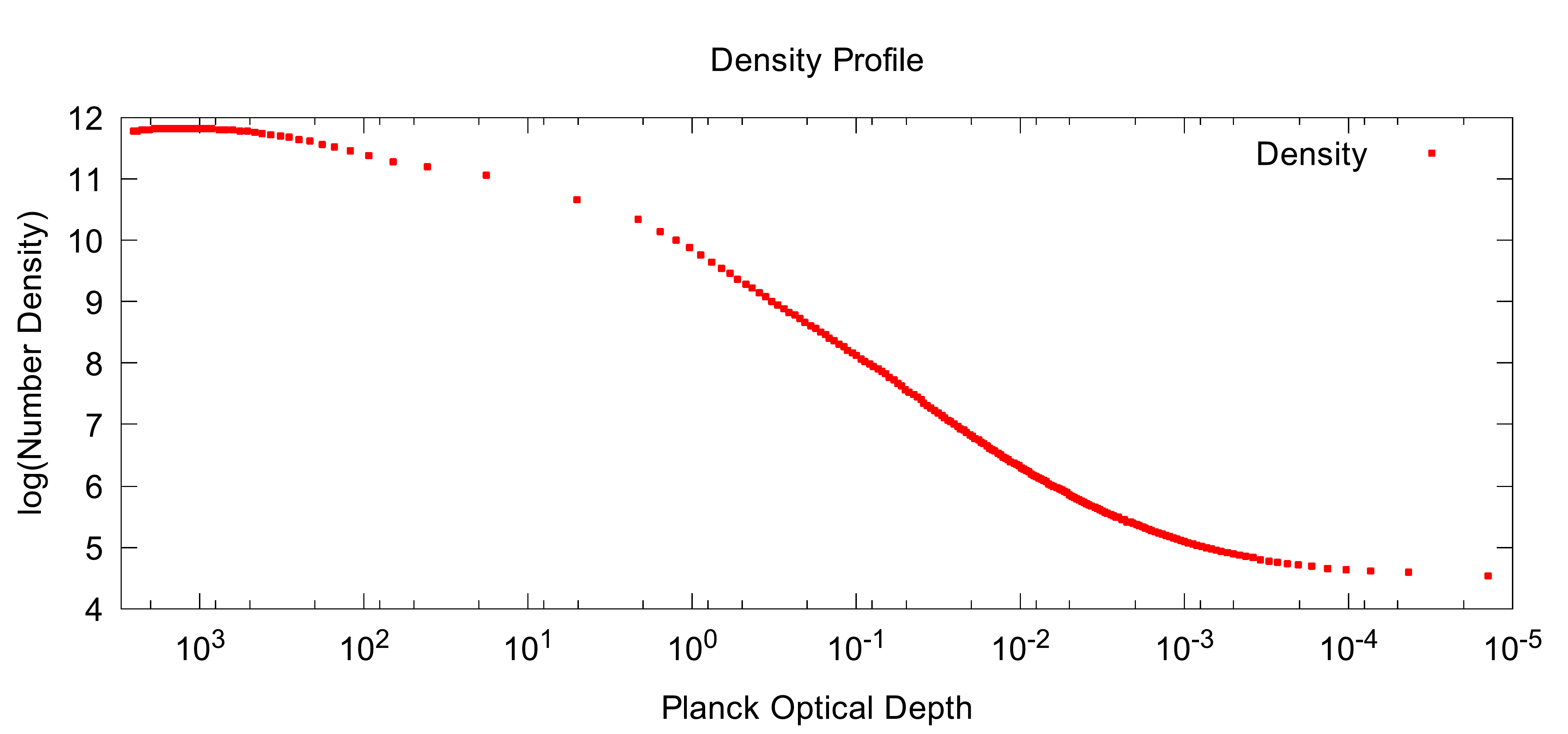}
		\plotone{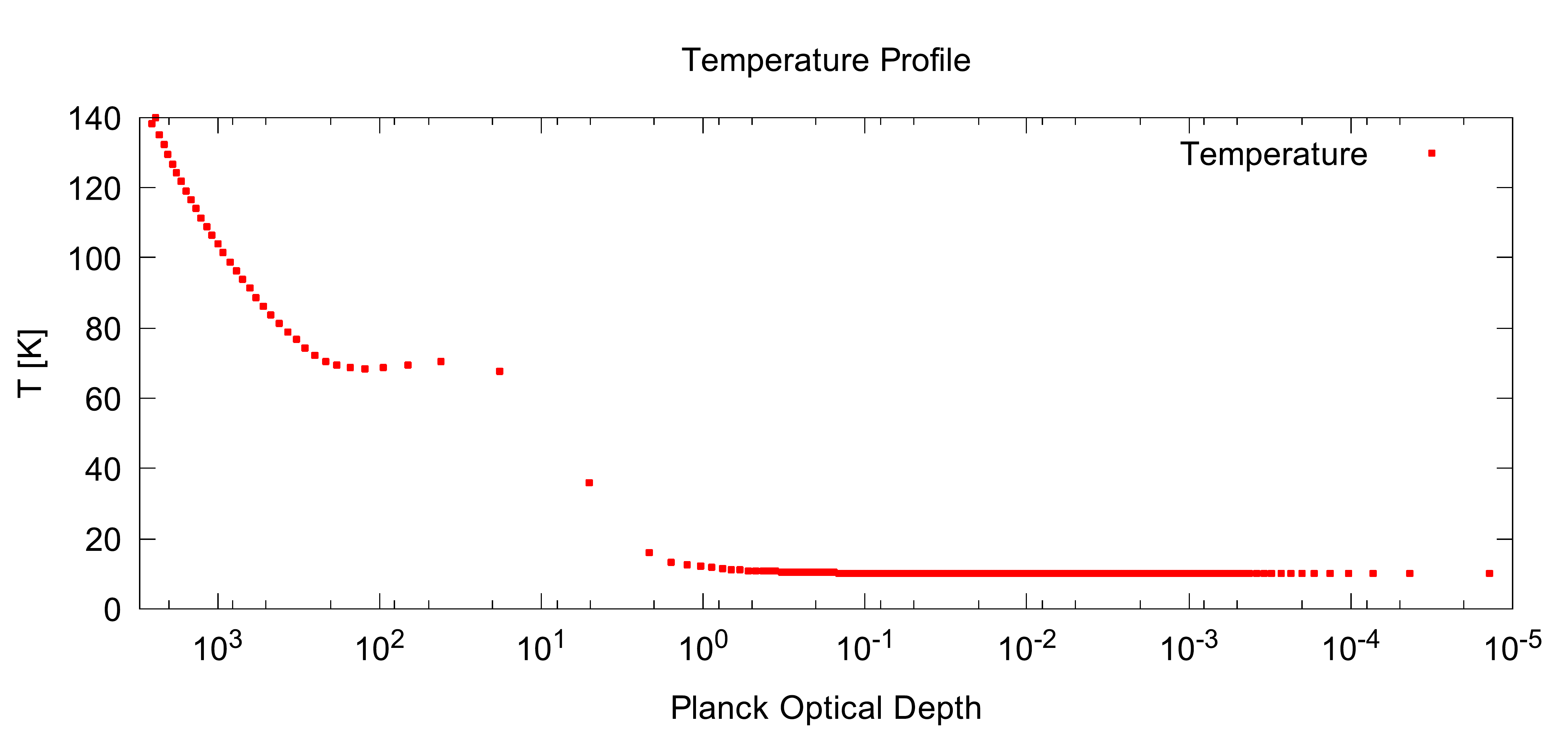}
	\caption{Density and temperature plotted against Planck optical depth, in the late stage of a simplified protostellar collapse simulation. A hot and dense interior region (the first hydrostatic core or Larson core) has formed, surrounded by a cold and thin envelope.}
	\label{fig:RhoundT}
\end{figure}

Figure \ref{fig:RhoundT} shows the profiles of density and temperature in the late stage of the collapse, when the total (Planck) optical depth $\tau$ from the core to the outside has reached 2500. 

At this point, we are not looking to provide a realistic simulation of protostellar collapse. We use optically thin cooling until the total optical depth reaches 0.1, and the constant-$\chi$-method with the gray approximation after that. Whenever the system's total (Planck) optical depth crosses certain thresholds ($\tau$=1, 5 , 1000 and 2500), we perform multiple calculations in a single time step (Constant-$\chi$-method as well as numerical angle integration with various resolutions). This provides snapshots of the difference in the results at these time steps.

Since no analytic solution for the heating is available, we have to evaluate the accuracy by considering the total radiative cooling (same as in section \ref{sec:ConstT}). It must be mentioned that this is not unproblematic, because the error in the total cooling is not necessarily representative of the error in the heating and, therefore, in the total radiation term. This is because the radiative exchange between shells $i$ and $j$ is proportional to $A_{\mathrm i \leftarrow \mathrm j} \left(S_{\mathrm j}-S_{\mathrm i}\right)$, so the total cooling is proportional to $S_{\mathrm i} \sum_j{A_{\mathrm i \leftarrow \mathrm j}}$, but the total heating is proportional to $\sum_j{A_{\mathrm i \leftarrow \mathrm j} S_{\mathrm j}}$. This means that, depending on the temperature of shell $j$, a given error in the calculation of $A_{\mathrm i \leftarrow \mathrm j}$ may have a much larger or much smaller effect on the heating than on the cooling. In consequence, the error in the total radiation term may be quite different from the error in the cooling alone. We will consider the total radiation term in section \ref{sec:Totals}, but in this section, we will only evaluate the relative error in the cooling.

\subsubsection{Constant-$\chi$-Method}
\label{sec:ConstChiTest}
\begin{figure}[tbp]
		\plotone{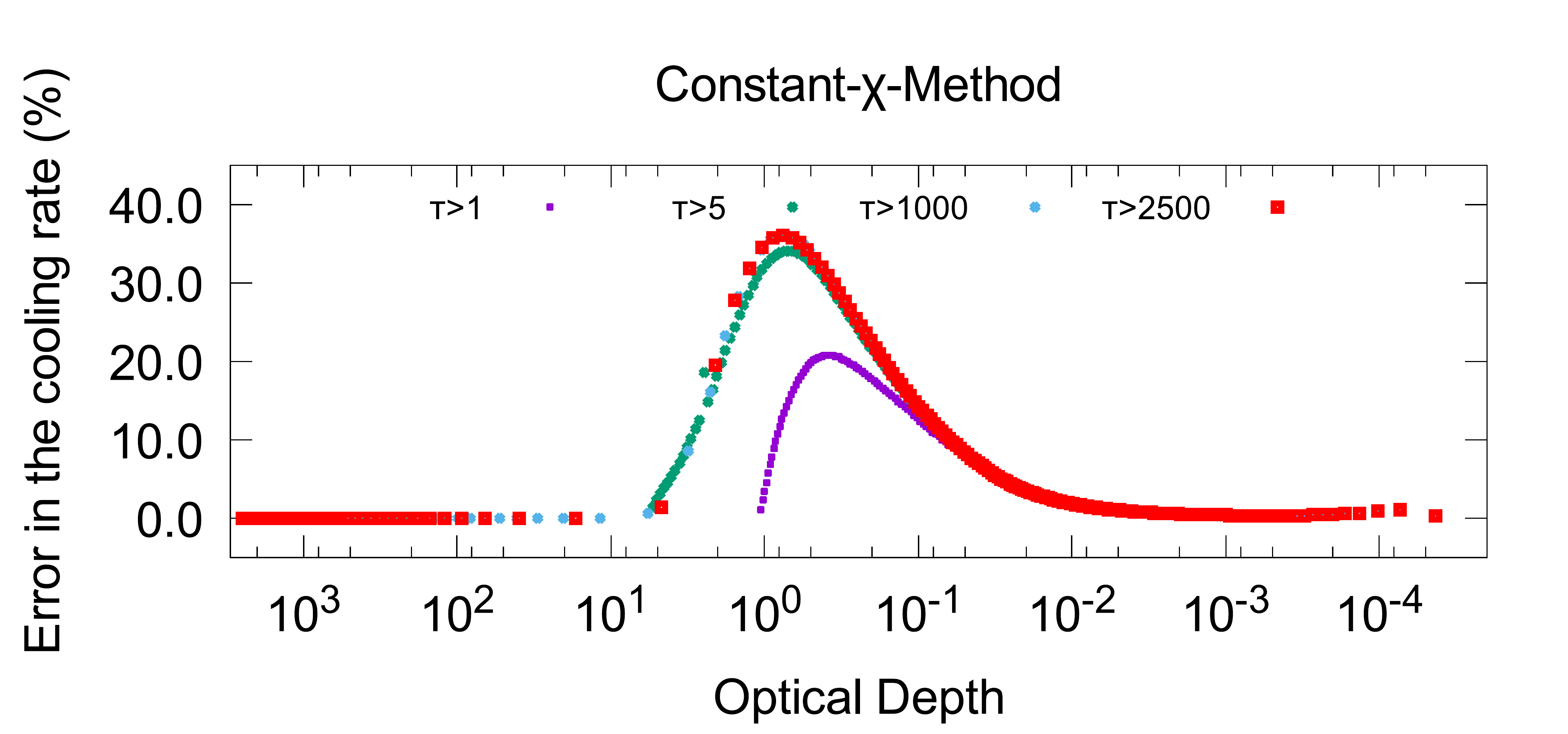}
	\caption{Relative error in the cooling rate for the constant-$\chi$-method at different stages in the collapse.}
	\label{fig:EI}
\end{figure}

Consider first the results for the constant-$\chi$-method (Figure \ref{fig:EI}). Recall that for this method, we took the average value of $\chi$ between $i$ and $j$ in the radial direction, and assumed that this same value can be used for every direction. Since in reality, $\chi$ near the center is now much larger than further outside, we expect this method to significantly overestimate the radiative exchange in the region where the external exchange is important. The plot confirms this: While the result is nearly exact close to the outer boundary, the overestimation increases towards the inside and reaches a maximum near the photosphere. After that, the error decreases again for several reasons: In the optically thick region, directions that are much different from the radial direction play a small role; near the center, $\chi$ does actually approach direction-independency due to symmetry; and for the very optically thick cases, the external term becomes negligible compared to the internal term.

The maximum error is about 20\% in the optically intermediate ($\tau=1$) case, but increases to 35\% when the core is fully optically thick. At this point, the core's optical thickness is essentially infinite so further increases have no more effect.

\subsubsection{Numerical Angle Integration}
\label{sec:NumTest}
\begin{figure}[htbp]
		\includegraphics[width=0.5\textwidth]{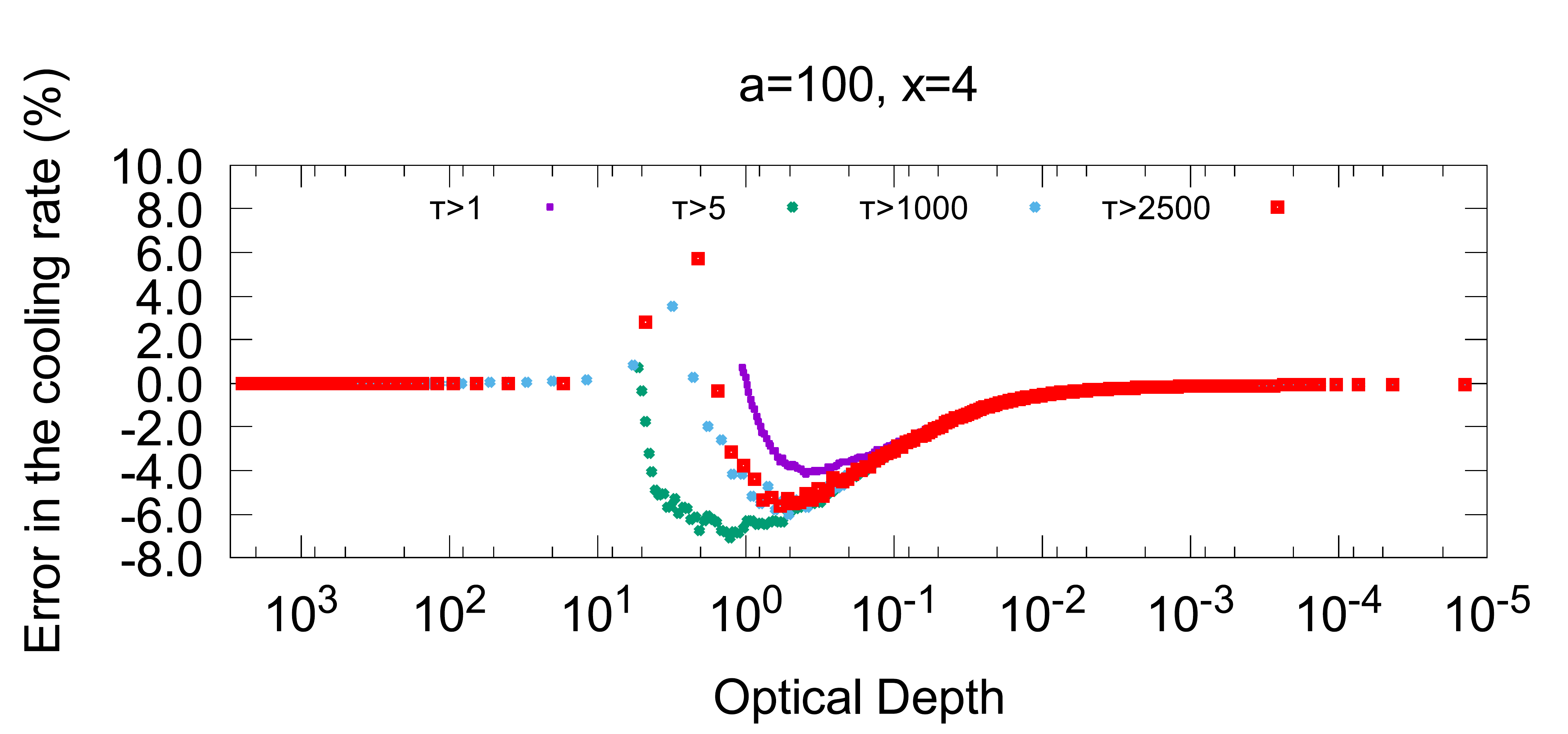}
		\includegraphics[width=0.5\textwidth]{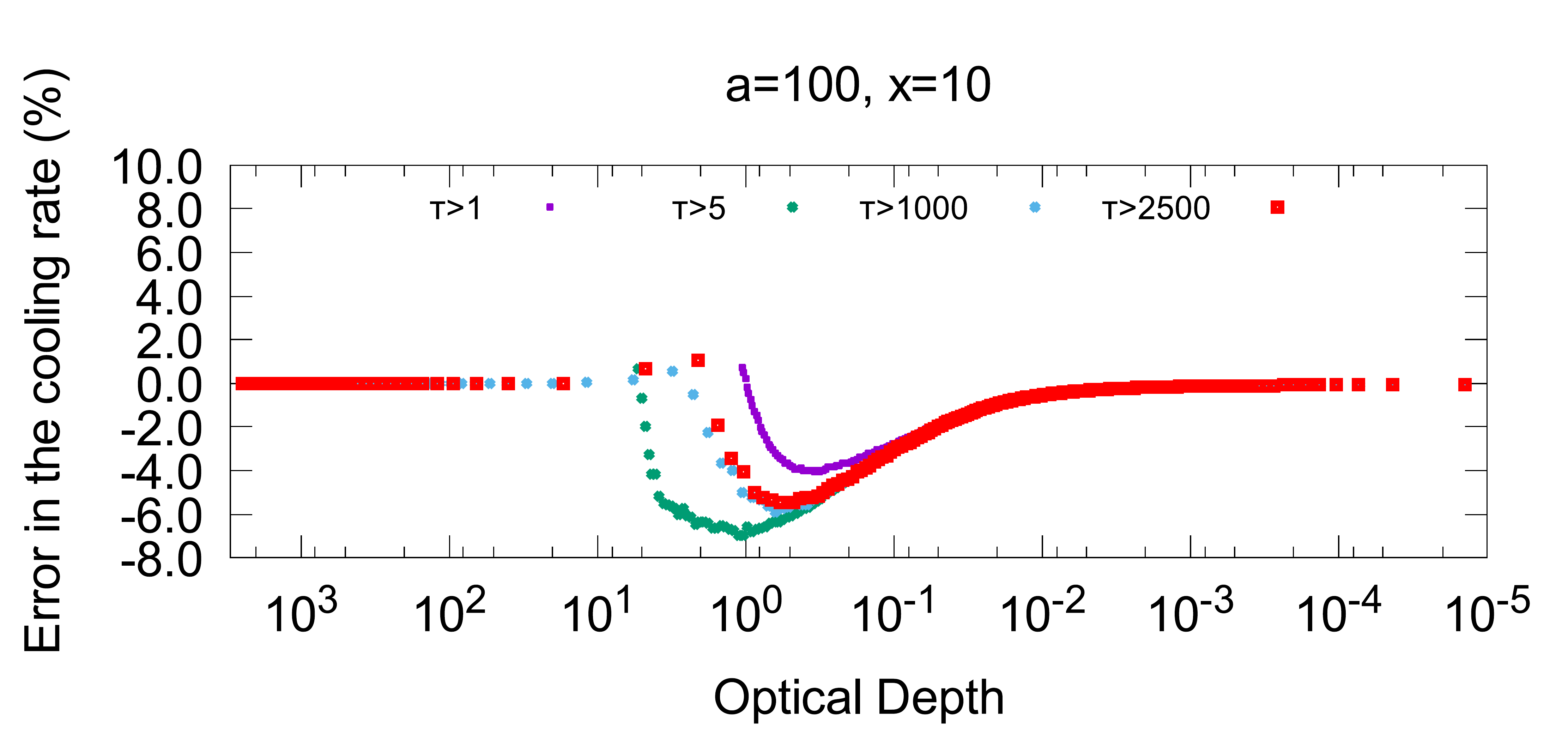}
		\includegraphics[width=0.5\textwidth]{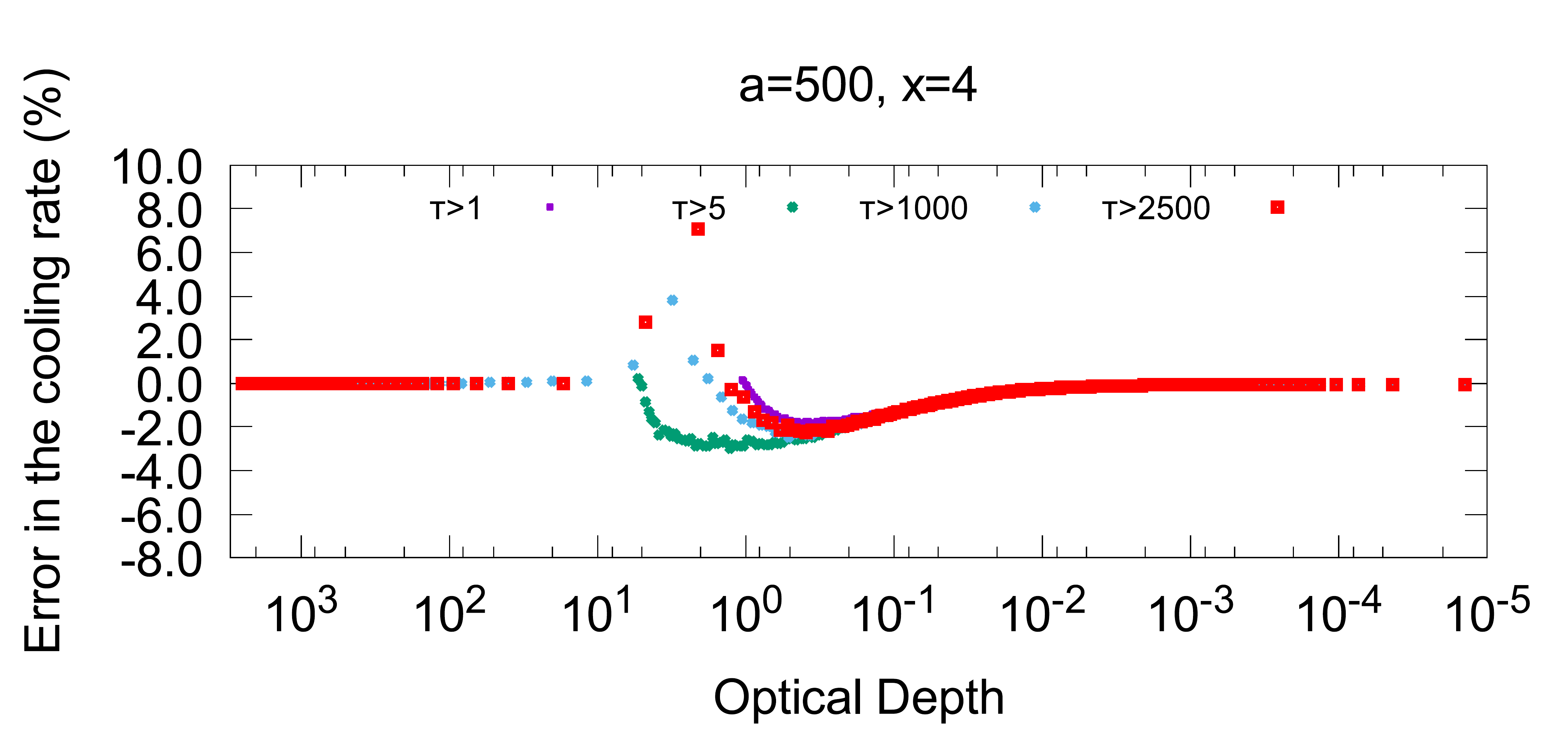}
	\includegraphics[width=0.5\textwidth]{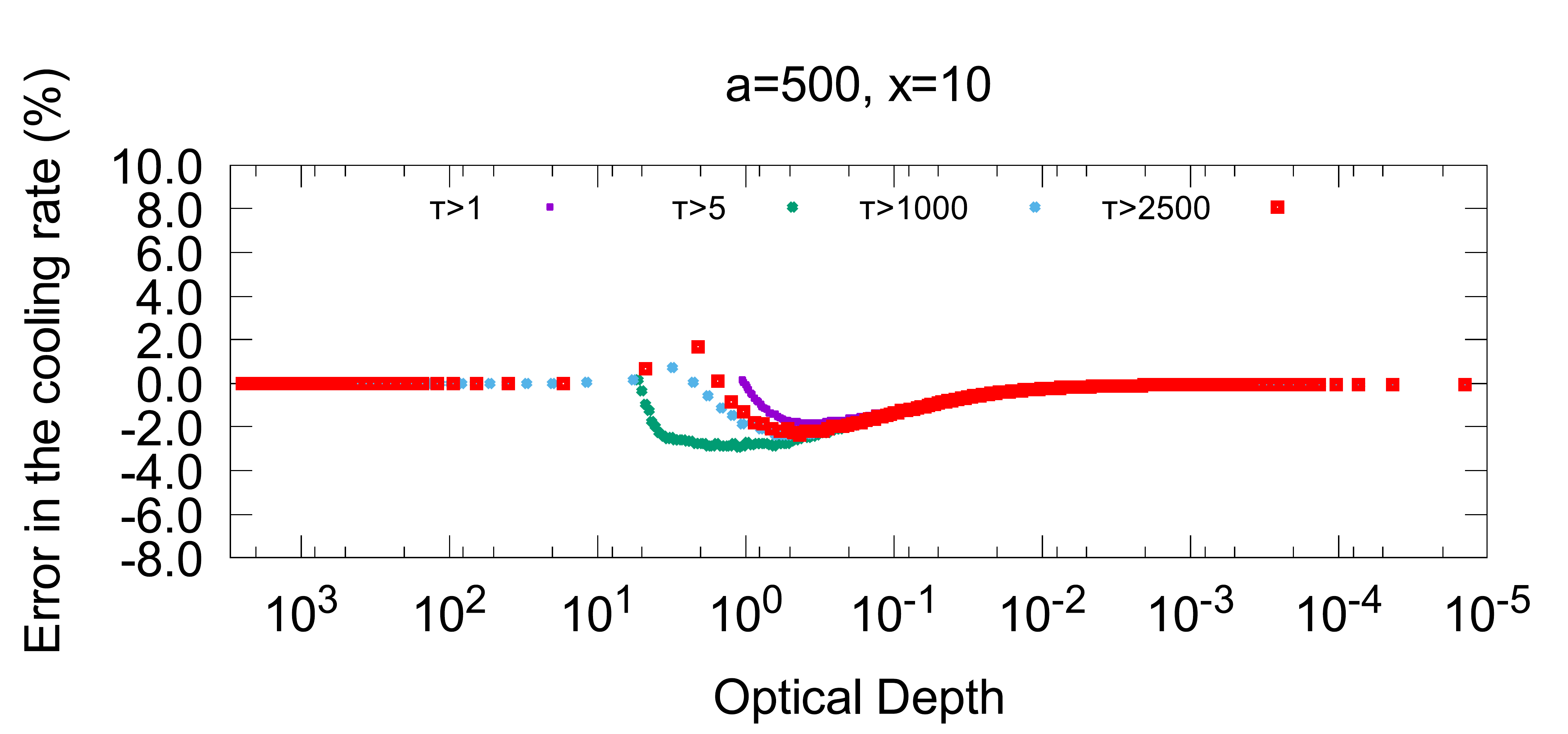}
		\caption{Relative error in the cooling rate for numerical angle integration at different stages in the collapse. Subshell resolutions $x=4$ and $x=10$, and angular resolutions $a=100$ and $a=500$ are shown.}
	\label{fig:NumTest}
\end{figure}

The error in the cooling rate at the same time steps as in section \ref{sec:ConstChiTest} is shown in Figure \ref{fig:NumTest}. 

As expected, there is no error in the far outer and deep inner regions. Near the photosphere, we see two different kinds of erroneous behavior: For all optical depths, there is an underestimation of the result reminiscent of Figure \ref{fig:ConstChiCooling} 's $\tau_{\mathrm S}=0.1$ case. The magnitude of this error depends on the angular resolution. In addition, for the very optically thick cases, there is also an overestimation behind the photosphere. The magnitude of this error depends on the subshell resolution.

The latter error naturally increases with the optical depth per shell, but reaches a maximum at $\tau_{\mathrm S} \approx 3$. At even larger values, the internal term starts to dominate the external term, so the error in the latter becomes irrelevant. The internal term does not suffer from this error since it integrates over the radius analytically.

\subsubsection{Total Radiation Term}
\label{sec:Totals}
Except for the thermal relaxation test (section \ref{sec:TRM}), up to this point, we have tried to evaluate our method by comparing the total radiative cooling to the analytic value. As was mentioned in section \ref{sec:VarChi}, unlike the radiative cooling, the error in the heating depends not only on the optical depth, but also on the temperature profile. This makes it difficult to predict how a given error in the cooling will translate to an error in the total.

\begin{figure}[htbp]
		\plotone{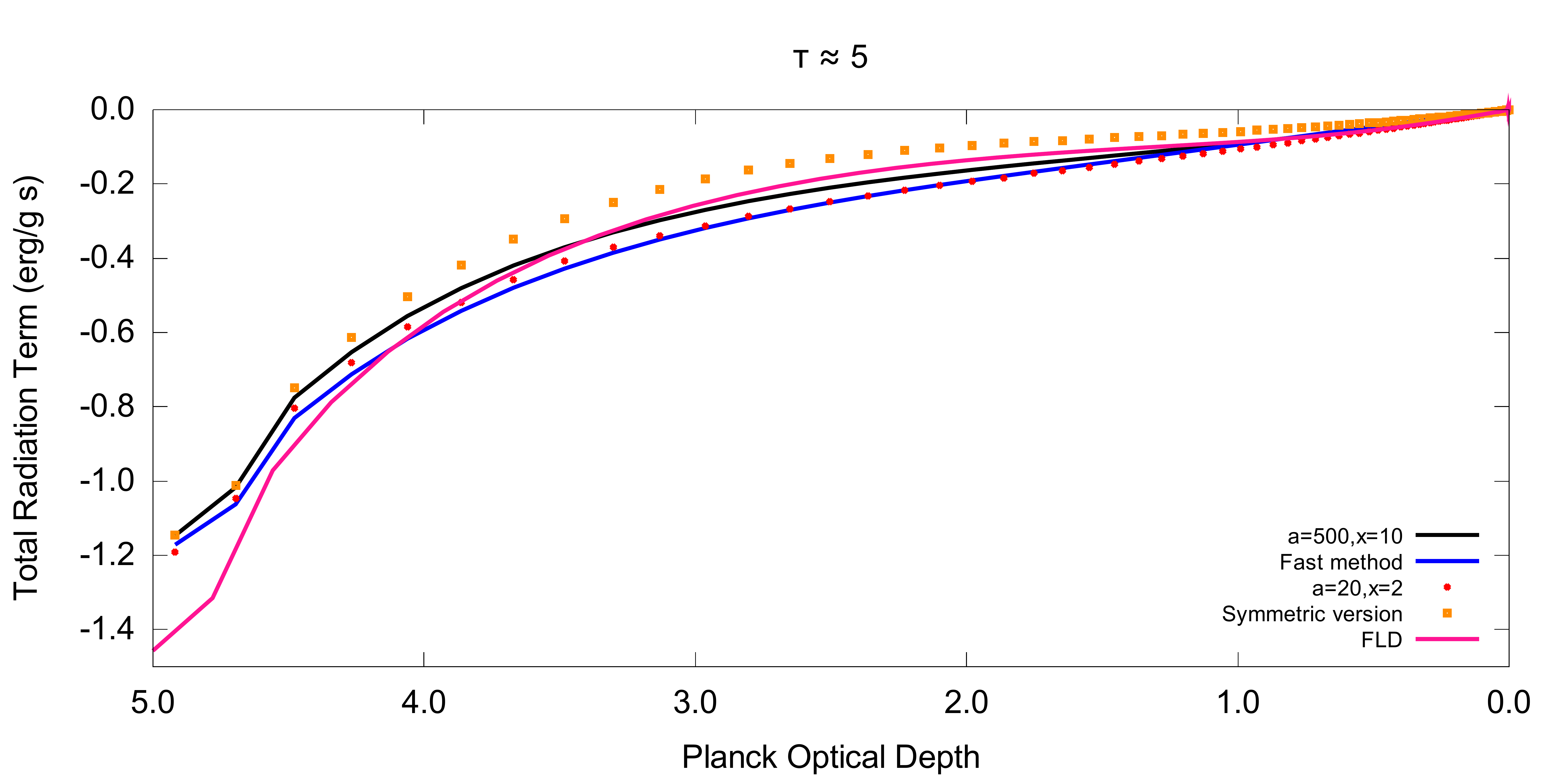}
		\plotone{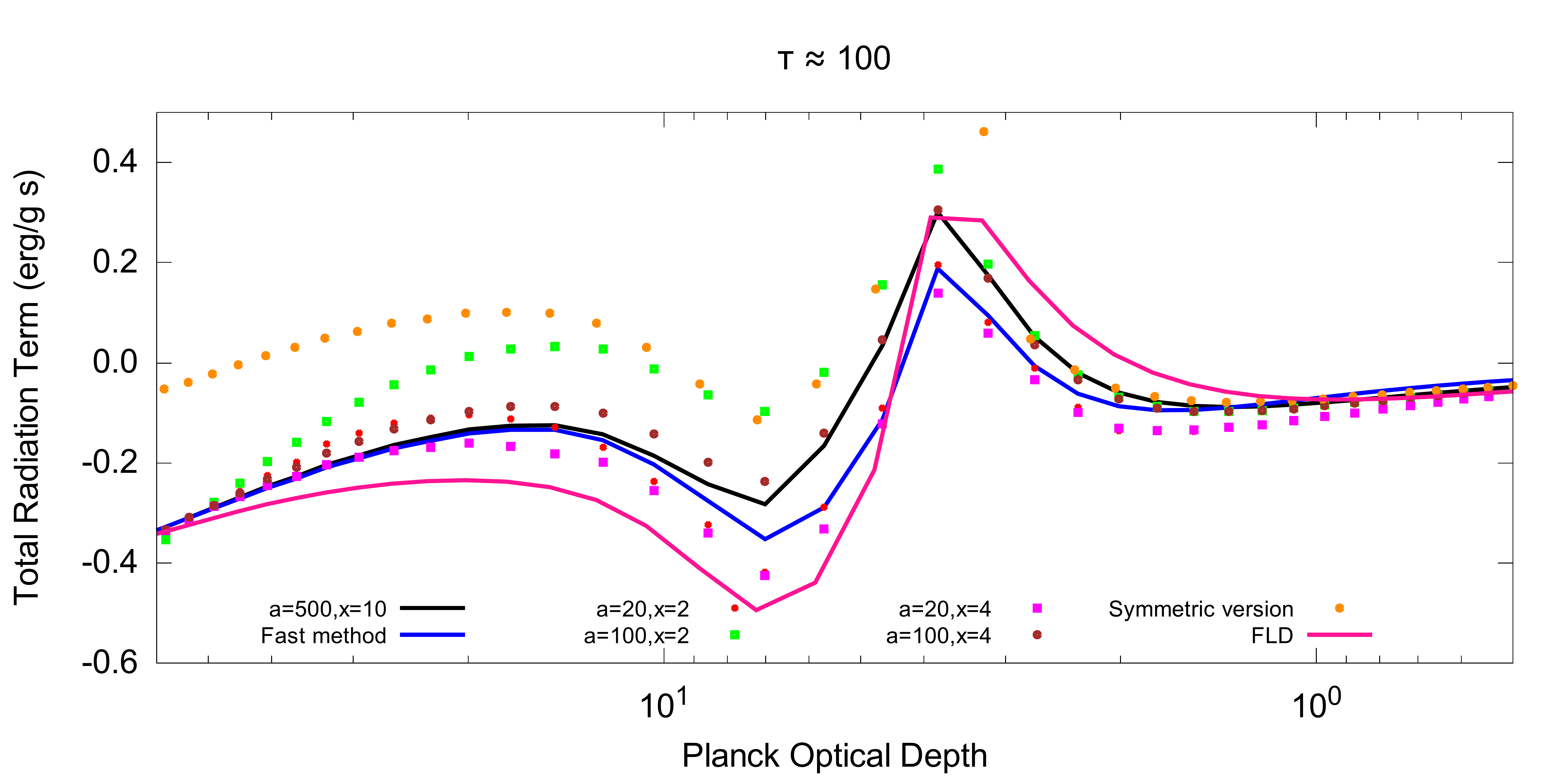}
		\plotone{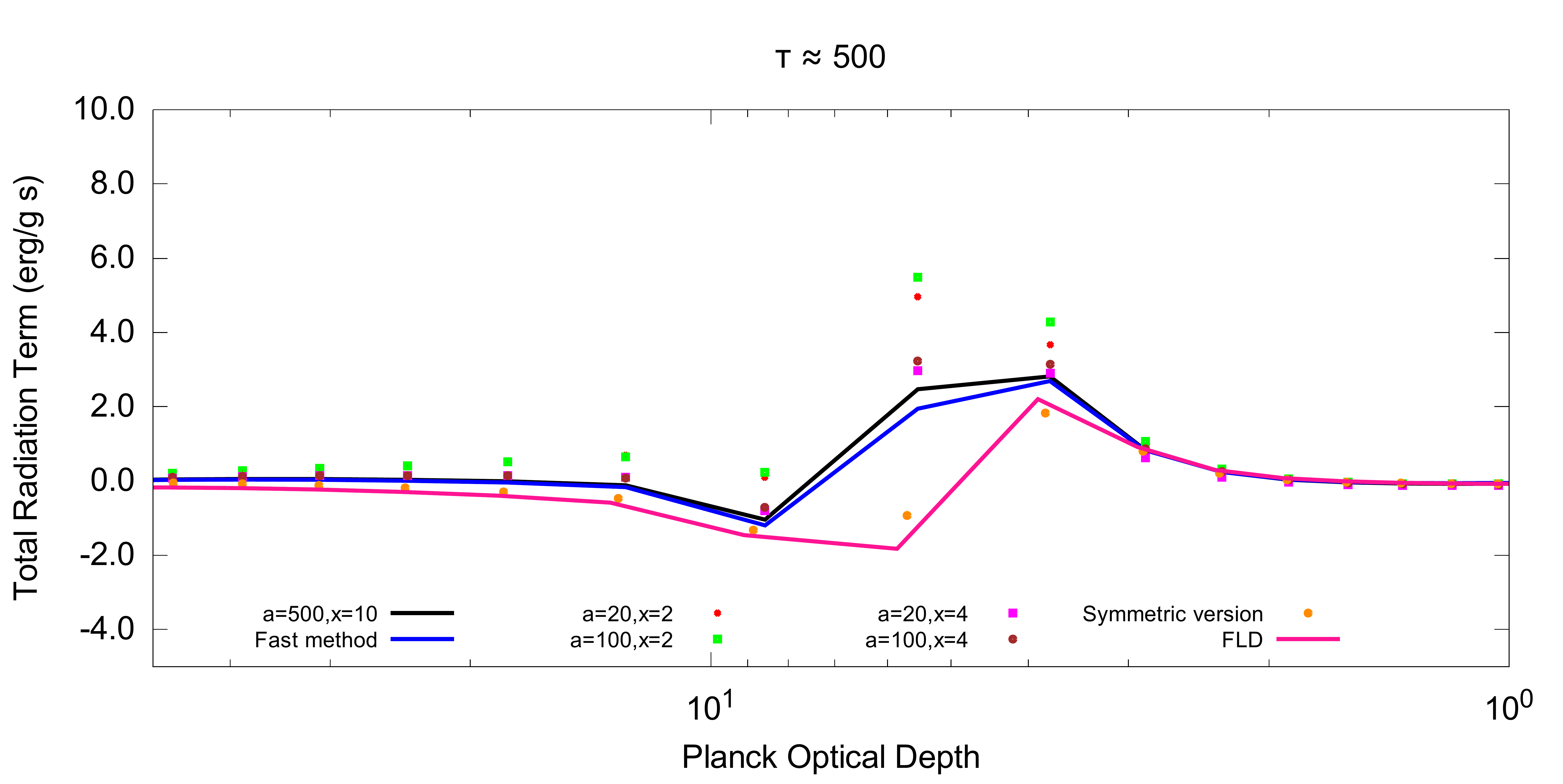}
	\caption{Total radiation terms for different resolutions at various stages during the collapse. Note that in the upper panel only, the $x$-axis is linear instead of logarithmic. The resolution for the symmetric version is a=500,x=10. The results for a flux-limited diffusion (FLD) calculation are also included for comparison.}
	\label{fig:Totals}
\end{figure}

Figure \ref{fig:Totals} compares the total radiation terms for different resolutions and for the constant-$\chi$-method at various stages in the collapse. We also include a high-resolution calculation for the symmetric version of the method discussed in section \ref{sec:NumImp}.

We concentrate on the photospheric region since the differences are due to the external term, which is negligible in the deep interior and far outside. We notice several tendencies:

\begin{itemize}
\item The net effect of radiation is a radially outward transfer of energy, as should be expected (net cooling inside, net heating outside).
\item The constant-$\chi$-method consistently shows a smaller value than the numerical integration. This is understandable since, as was mentioned briefly in section \ref{sec:ConstChi}, the overestimation of $A_{\mathrm i \leftarrow \mathrm j}$ is smaller for $i>j$ than for $i<j$. This leads to the relative overestimation in the heating being smaller than in the cooling, since the former comes mostly from the inside.
\item As the total optical depth increases, so does the jump in $\chi$ at the boundary of the first core, which becomes more and more sharply defined. Because of this, the number of shells where the external term is relevant continues to decrease. This is advantageous for the computational speed; however, if a higher resolution of the (post-)photospheric region is desired, some kind of mesh refinement method is needed.
\item The differences are most noticeable in the $\tau \approx 100$ case, when there is an optically thick region where individual shells are still optically thin. The error in the fast method is more significant here. Because of this, the pure constant-$\chi$-method may not be a good choice for a realistic calculation.
\item Concerning the resolution of the numerical integration: Similar to the results in Figure \ref{fig:NumTest}, we see that low angular resolution leads to an overall underestimation, while low subshell resolution leads to an overestimation in the optically thick region but has almost no effect further outside. The behaviour of the fast method essentially corresponds to a calculation with minimal angular and infinite subshell resolution. The differences between $a=100$,$x=4$ and $a=500$,$x=10$ are small, showing that the method has converged.
\item The use of the symmetric definition of $\ix{A}{ij}$ has only a small effect for optically thin shells, since there is little difference then between fixing the position in shell $i$ and integrating over shell $j$ or doing it the other way around. It also does not matter for very optically thick shells, where the internal term dominates. However, for the intermediate case (interior region of the middle panel in Figure \ref{fig:Totals}) the result is significantly different from the non-symmetric case, despite using the same angular and subshell resolutions. Note that the differences here do not necessarily mean that the symmetric version is less accurate; it is rather that the two versions attempt to do different things. The non-symmetric version is designed to calculate the radiation term at the shell center as accurately as possible, while the symmetric version is intended to calculate a shell-average that guarantees energy conservation.
\item As should be expected, the flux-limited diffusion approximation causes significant errors in the photospheric region where individual shells are not yet optically thick.
\end{itemize}

These results are specific for this particular optical depth and temperature profile, so when applying the method to a different system, the results may differ as well. However, the general situation of a hot, optically thick interior surrounded by a cold, thin envelope is a rather common state in astrophysics.

The differences between the symmetric and non-symmetric versions of the method are expected to disappear in the limit of optically thin shells, since in that case the variation of the radiation term within a shell should vanish. In principle, we could confirm this by performing a protostellar collapse calculation with sufficiently high resolution to keep all shells optically thin throughout the simulation. In practice, the computational cost makes this impossible: The system optical depth after first core formation is already larger than $10^3$, so the required number of shells would be $10^4$ at the very least, and $10^5$ for truly optically thin shells. The presence of optically thick shells is not an issue since they are dominated by the internal term, which is the same for both variants of the methods. However, there will always be a region in which shells are neither fully thin nor fully thick. This is the problematic region in which the external term is important. Increasing the resolution will not make this region disappear, but only move it closer to the system center. 

Therefore, we test the convergence of the two variants by using a simplified system in which the density is constant, the velocity is zero and all hydrodynamic calculations are switched off. We introduce a temperature gradient (temperature decreasing outwards from the center) to create a situation similar to the realistic simulation, and the (gray) opacity increases quadratically with temperature. The latter is important to introduce some asymmetry between the different shells: If density, opacity, and radial size of all shells are the same, then the asymmetry that causes the differences between the two variants of the method (section \ref{sec:General}) disappears, and both variants give identical results irrespective of the temperature profile.

\begin{figure}[htbp]
		\plotone{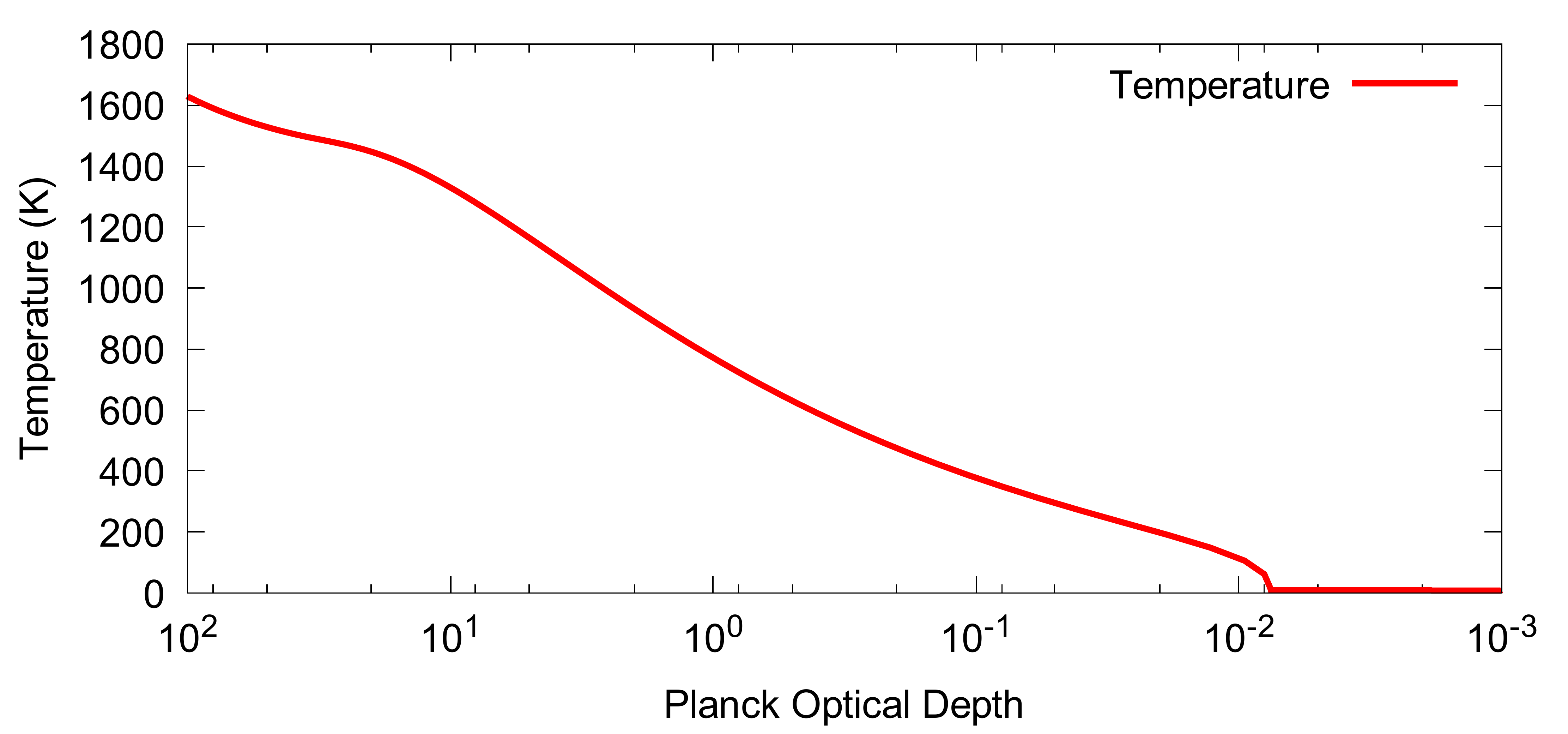}
		\plotone{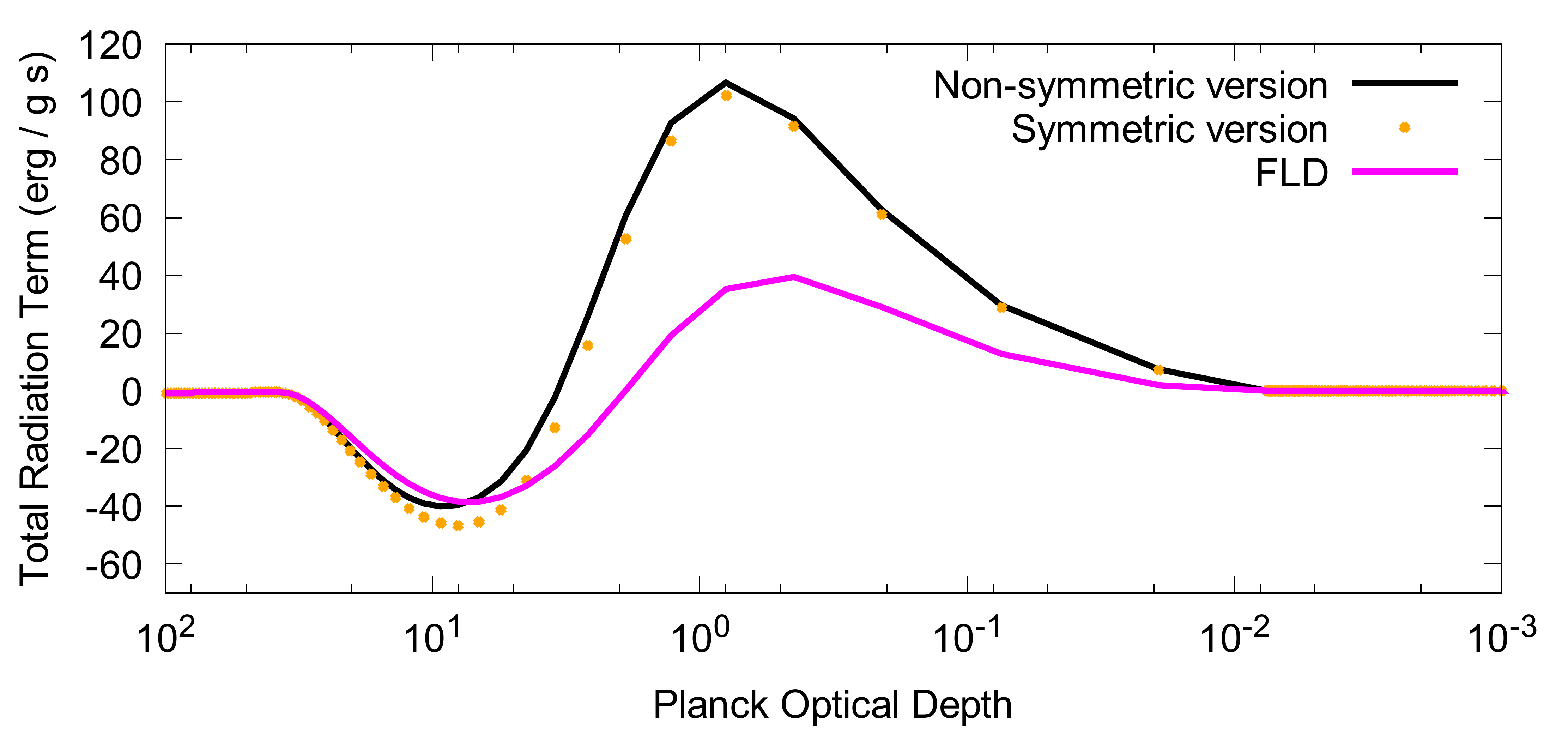}
		\plotone{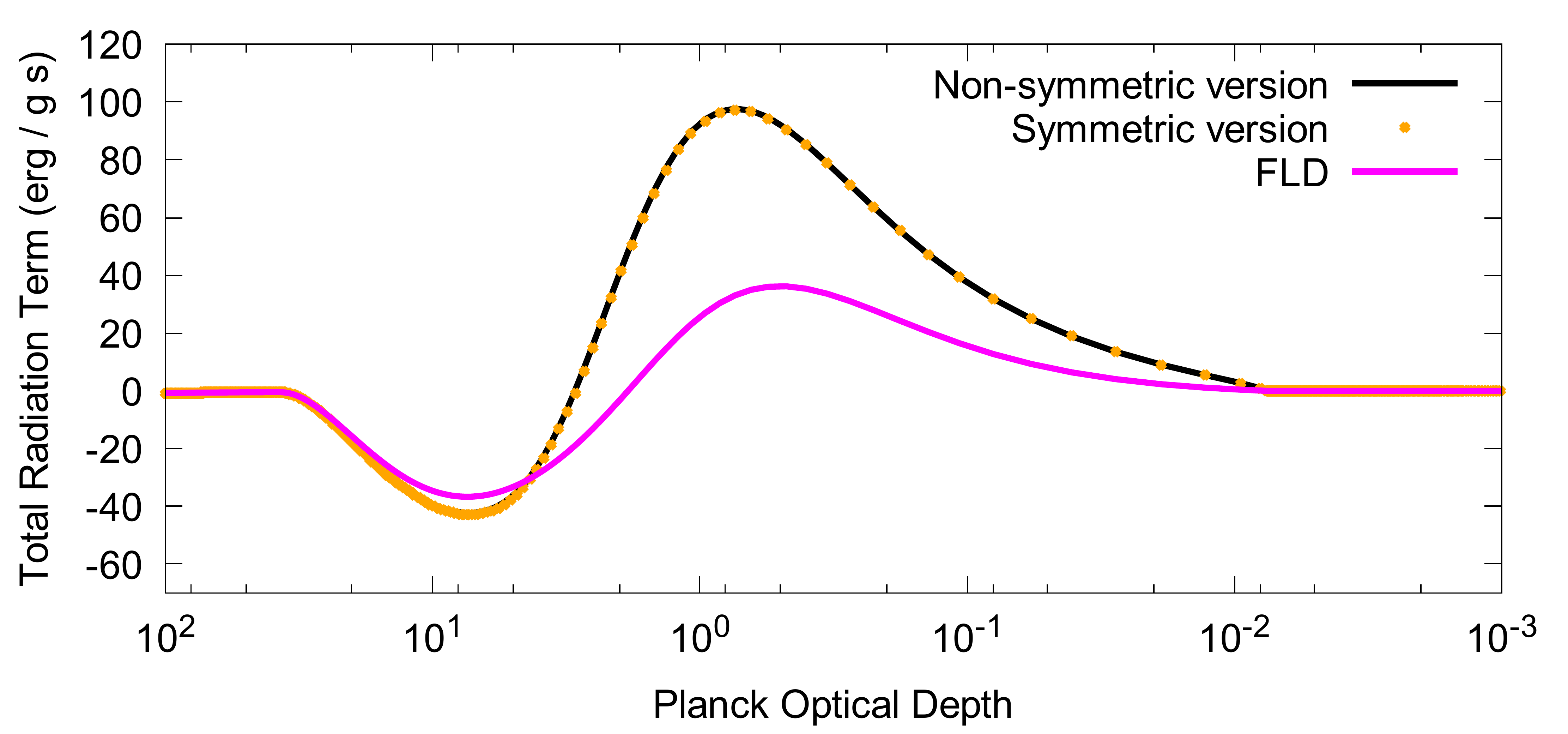}
	\caption{Convergence test for the symmetric and non-symmetric versions of the method. The upper panel shows the temperature profile used for this test. The number of shells in the middle panel is 250, and the optical depth per shell is close to 1 near $\tau=10^1$ and decreases further outwards. In the lower panel, the number of shells is 1000 and the optical depth per shell at $10^{1}$ is around 0.2, i.e. the shells are optically thin throughout the relevant region.}
	\label{fig:reftest}
\end{figure}

Figure \ref{fig:reftest} clearly shows that the two variants, whose results differ substantially in the low-resolution case (optically intermediate shells), are converged for sufficiently high resolution (optically thin shells). Interestingly, the high-resolution results are closer to the low-resolution results of the symmetric than of the non-symmetric method, which indicates that, at least for this setup, the former may be preferable. The flux-limited diffusion results are also included, but as should be expected in this region of optically thin/intermediate shells, they are quite different from the results of our method.

\subsubsection{Combined Method}
\label{sec:Combo}
In section \ref{sec:Totals}, we found that the constant-$\chi$-method is too inaccurate to be used for the whole calculation, at least during the time when there is an optically thick region with optically thin shells. The numerical angle integration is much more accurate, but also much slower. In this section, we will attempt to combine the two in order to arrive at a compromise between speed and accuracy.

\begin{figure}[htbp]
		\plotone{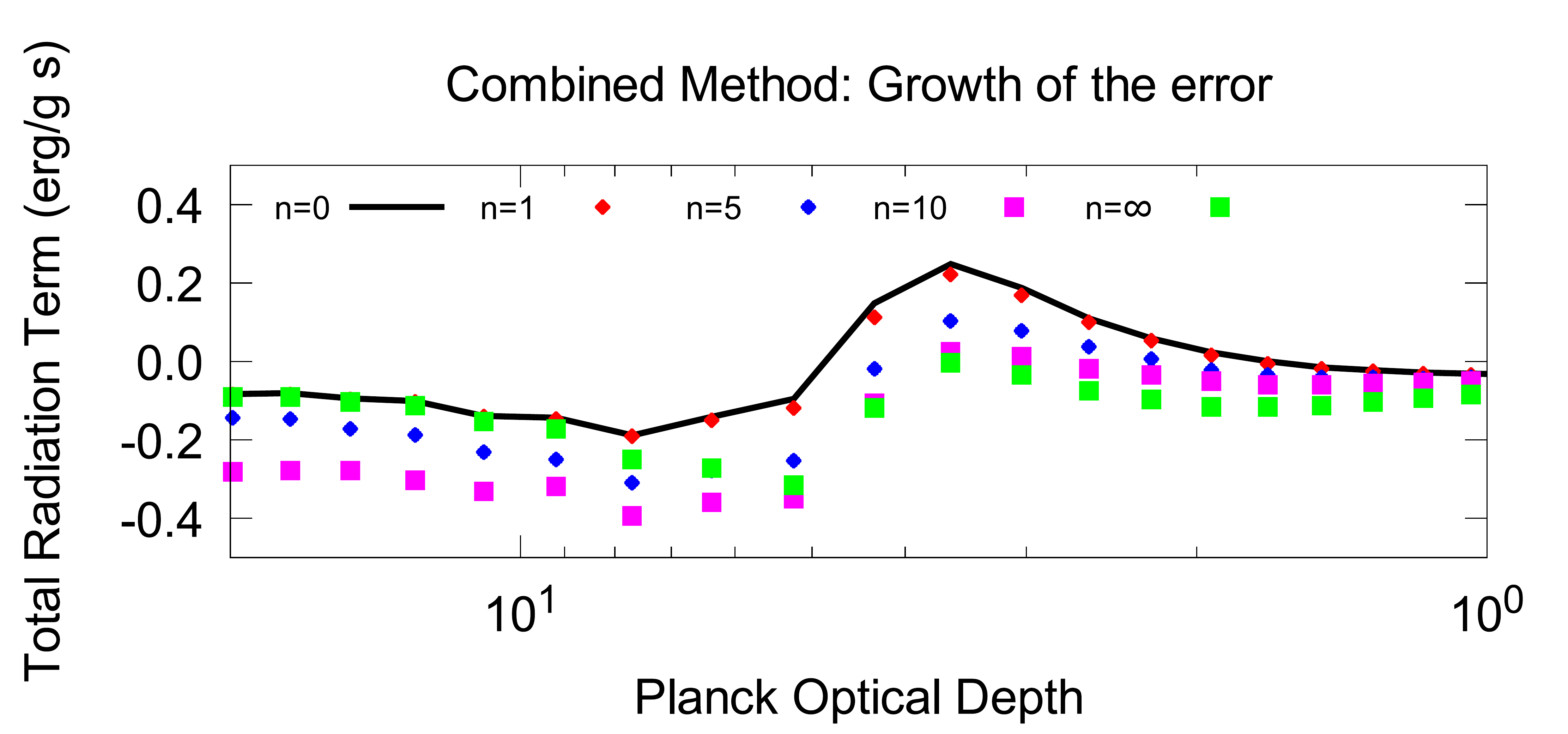}
	\caption{Combined method. Snapshots of the photospheric region at a time step when $\tau_{\mathrm P} \approx 100$, equivalent to the middle panel of Figure \ref{fig:Totals}. $n$ indicates the number of time steps since the correction factors $C_{\mathrm i \leftarrow j}$ were last calibrated.}
	\label{fig:Combo}
\end{figure}

This can be done in the following way: At time step t, both slow and fast methods are used to calculate the $A_{\mathrm i \leftarrow \mathrm j}$. Then, the ratio of the two is written into an array of correction factors $C_{\mathrm i \leftarrow \mathrm j}=A_{\mathrm i \leftarrow \mathrm j}\left(fast\right) / A_{\mathrm i \leftarrow \mathrm j}\left(slow\right)$. For the following time steps, only the fast method is used, and the resulting $A_{\mathrm i \leftarrow \mathrm j}$ are divided by $C_{\mathrm i \leftarrow \mathrm j}$. After $n$ time steps, both methods are used and the correction factors recalibrated. From Figure \ref{fig:Combo}, we can see that after 5 steps, the results have already become quite inaccurate, so this combined method could only reduce the computational cost by a factor of a few at most. However, this is not true over the whole calculation: Figure \ref{fig:Combo} represents the worst case, as it is during the transition phase when the optical depth structure is changing quickly. In earlier and later stages, acceptable values of $n$ can be much larger. Therefore, instead of being constant, $n$ should be changed dynamically during runtime. We suggest the following method:

First, we choose a number $m$ (for example, $m=0$ initially) and set $n=2^m$. Every $n$ timesteps, we perform both slow and fast calculations as described above to obtain the radiation terms ${dU / dt}_{\mathrm {fast}}$ and ${dU / dt}_{\mathrm {slow}}$. We then calculate the relative error in the internal energy that would result after $n$ timesteps:

\begin{equation}
\delta U = \frac{\abs {{dU / dt}_{\mathrm {fast}} - {dU / dt}_{\mathrm {slow}}} n \Delta t} {U}.
\end{equation}
We evaluate this for every shell, and if it is larger than some critical value in any shell, we reduce $m$ by 1. If it is smaller than some other critical value in every shell, we increase $m$ by 1. In this way, the frequency of slow and accurate radiative calculations changes dynamically during the calculation.
This method still risks having large errors in some timesteps when $m$ is too large before it can adapt. We can avoid this by including a rollback function as a safety: Every time we perform a slow calculation, we also store all the basic variable arrays such as density, temperature etc. If the error in the next slow timestep ($n$ timesteps later) is too large, in addition to reducing $m$, we also jump back to the stored values. This guarantees that the maximum error is always smaller than some critical value defined by the user.
In practice, this method succeeds in significantly increasing the computational speed compared to the pure slow method, since numerical calculations are only very rarely needed after the first core has established itself and the optical depth structure no longer changes quickly.

\subsubsection{Computational Speed and Optimization}
\label{sec:Opt}

Table \ref{tab:comptime} shows the wall-clock time for calculations using the pure fast or pure slow methods. In our calculation, there are three different constraints on the timestep: 

\begin{itemize}
\item The basic hydrodynamical timestep, determined by the condition that neighboring shells must not overtake each other, i.e. $\Delta t = x \Delta r / (\Delta v + \ix{c}{s})$, where $\Delta r$ and $\Delta v $ are the differences in radius and velocity of two neighboring shells, and x is a number between 0 and 1. Here we choose x=0.1.
\item The radiation timestep. Since the radiation term is very sensitive to temperature, it must be recalculated after the temperature in any shell has changed by more than a certain percentage. Here we choose 1 percent. Note that the exchange coefficients $\ix{A}{ij}$ are not updated at this point.
\item The opacity timestep. The $\ix{A}{ij}$ depend on the optical depth structure of the system, which in turn is determined mostly by the density structure. We update these coefficients whenever the density in any shell has changed by more than 1 percent since the last update. Whenever this is done, the radiation term is also recalculated.
\end{itemize}

The purely hydrodynamical timestep requires much less computational effort than the ones related to the radiative transfer, as evidenced by the fact that the calculation without RT takes less than a minute. The radiation timestep and especially the opacity timestep are far more time-consuming, the latter even more so if numerical angle integration is used. The calculations shown in Table \ref{tab:comptime} required roughly 7500-10000 hydrodynamical timesteps (depending on the number of shells), 3000 radiation timesteps without $\ix{A}{ij}$ updates, and 1500-2000 opacity timesteps. Note that the ratios of these numbers may be vastly different in other systems, depending on the time scales at play. The flux-limited diffusion method is comparable in speed to the fast method at 200 shells, but scales better with higher resolutions since it is $O\left(f n\right)$. Its speed depends on how many steps are allowed for the radiation field to converge, however.

\begin{table}
\centering
\begin{tabular}{c | c | c | c | c}
Method Type & Shells & Angles & Subshells & Wall-clock time (minutes) \\
\hline
No rad. transfer & 200 & - & - & $<$ 0.5 \\
Fast & 200 & - &  - & 6\\
Fast & 400 & - & - & 21\\
Slow & 200 & 20 & 2 & 17\\
Slow & 200 & 100 & 2 & 61\\
Slow & 200 & 20 & 4 & 27\\
Slow & 400 & 20 & 2 & 71 \\
FLD & 200 & - & - & 9 \\
FLD & 400 & - & - & 17\\
\end{tabular}
\caption{Simplified gray collapse calculations for various shell, angle, and subshell resolutions. The tests were run on a single 3.4 GHz processor and stopped once the total optical depth reached 1500.}
\end{table}
\label{tab:comptime}

As mentioned before, the external term is by far the slowest of the three terms in the radiative calculation. If the number of shells is $n$ and the number of frequencies considered is $f$, the fast method's external term scales as $O\left(f n^2\right)$, while the numerical integration over $a$  angles and with $x$ subshells scales as $O\left(a x f n^2\right)$, which for reasonable resolutions is two to three orders of magnitude slower. This is why optimization requires minimizing the number of numerical integrations, as we did in the previous section.

The separation of the computationally expensive external term from the other two, and the fact that it can be ignored in many regions, is of primary importance. In order to optimize speed, it is imperative to neglect the external term wherever possible, especially in the slow method. In our collapse simulations, we may ignore the external term in the far outer regions, and after first core formation also in the interior because individual shells there are optically thick. In addition, there may be regions where the external term is not negligible, but high accuracy is not required. In this case, instead of completely neglecting it, one may calculate it using only the fast method. These optimizations can reduce the actual time required by an order of magnitude or more compared to the values seen in Table \ref{tab:comptime}. In any case, the optimization strategy must be adapted according to the problem.

\section{Conclusion and Future Development}
In this paper, we have developed a new method of treating the radiative transfer in a spherically symmetric fluid system. Based on the transfer equation, we derived a scheme that is reduced to the calculation of exchange coefficients $A_{\mathrm i \leftarrow j}$ between pairs of shells (``external term''), between a shell and the radiation coming from outside the computational domain (``boundary term''), and between the shell center and other points within the same shell (``internal term''). The external term is by far the most problematic of these, since the calculations of the other two are both much faster and produce (nearly) exact results.

Numerical integration over all angles can solve the external term to a high degree of accuracy if a sufficiently large number of angles is used, however it is quite slow. By assuming a direction-independent absorption coefficient, the speed can be increased by two to three orders of magnitude, but at the cost of reduced accuracy. The resulting error is most evident in regions that are optically thick while individual shells are still optically thin. A combined method of numerical angle integration and constant-$\chi$-approximation can be used as a compromise between speed and accuracy.

Future improvements of the method will likely focus on speed optimization, either directly by increasing the speed of the numerical angle integration, or indirectly by increasing the accuracy of the fast method to make it more usable. Several approaches are possible:

\begin{itemize}
\item Flexible resolutions: In this paper, we have taken the angular and subshell resolutions to be constant throughout a given test run. It may be far more effective to change them during runtime. For example, $x$ only needs to be larger than 1 for shells in a certain optical depth range between about 0.1 and a few, and $a$ can be smaller near the center, where there is little exchange with shells smaller than $i$. 
\item For the constant-$\chi$-method, we use the average $\chi$ between shells $i$ and $j$ in direction $\tilde \theta = 0$. There may be better options, such as a weighted average between the $\tilde \theta=0$ and $\tilde \theta = \pi$ directions. In addition, the effect of the optically thick core could be approximated by defining a ``shadow region'' in the interior and reducing $A_{\mathrm {i} \leftarrow {j}}$ accordingly.
\item The main reason that the numerical integration is so slow is because of the high angular resolution required, which is a consequence of the fact that many rays emanating from a given shell $i$ fail to interact with shell $j$ if the latter is much smaller than $i$ (case 3 in Figure \ref{fig:Num}). However, instead of calculating $A_{\mathrm {i} \leftarrow {j}}$ and $A_{\mathrm {j} \leftarrow {i}}$ separately, one may calculate $A_{\mathrm {i} \leftarrow {j}}$ only for the case of $i<j$ and reuse this value when updating shell $j$. Aside from the obvious decrease by a factor of 2 in the computation time, this also reduces the angular resolution required, and has the additional advantage of guaranteeing energy conservation. But this method is problematic: $A_{\mathrm {i} \leftarrow {j}}$ is the exchange coefficient between $r_{\mathrm i}$ and the whole of shell $j$. As we discussed in section \ref{sec:NumImp}, this is conceptually different from $A_{\mathrm {j} \leftarrow {i}}$, the exchange coefficient between $r_{\mathrm j}$ and the whole of shell $i$. If one is willing to accept this inconsistency, a significant performance increase may be possible.
\end{itemize}

In conclusion, the performance of the method will depend strongly on steps taken for optimization, and which methods of optimization are effective will depend on the optical depth and temperature structure of the system in question.


\begin{thebibliography}{}

\bibitem[Bodenheimer et al. (1990)]{Bodenheimer1990} Bodenheimer, P., Yorke, H. W., Rozyczka, M., \& Tohline, J. E. 1990, \apj, 355, 651-660
\bibitem[Buchler (1979)]{Buchler1979} Buchler, J. R. 1979, J. Quant. Spectrosc. Radiat. Transfer, 22(3), 293-300
\bibitem[Castor (1972)]{Castor1972} Castor, J. I. 1972, \apj, 178, 779-792
\bibitem[Colella \& Woodward (1984)]{Colella1984} Colella, P., \& Woodward, P. R. 1984, J. Comput. Phys., 54(1), 174-201
\bibitem[Commer\c{c}on et al. (2010)]{Commercon2010} Commer\c{c}on, B., Hennebelle, P., Audit, E., Chabrier, G., Teyssier, R. 2010, A\&A,510, L3
\bibitem[Commer\c{c}on et al. (2012)]{Commercon2012a} Commer\c{c}on, B., Launhardt, R., Dullemond, C., Henning, T. 2012a, A\&A, 545,A98
\bibitem[Dzyurkevich et al. (2016)]{Dzyurkevich2016} Dzyurkevich, N., Commer\c{c}on, B., Lesaffre, P., \& Semenov, D. 2016, arXiv:1605.08032
\bibitem[Federrath (2015)]{Federrath2015} Federrath, C. 2015, MNRAS, 450, 4035
\bibitem[Frimann, J\o rgensen, \& Haugb\o lle (2016)]{Frimann2016} Frimann, S., J\o rgensen, J., Haugb\o lle, T. 2016a, A\&A, 587, A59
\bibitem[Gonz\'alez et al. (2015)]{Gonzalez2015} Gonz\'alez, M., Vaytet, N., Commer\c{c}on, B., Masson, J. 2015, A\&A, 578, A12
\bibitem[Hennebelle \& Charbonnel(2013)]{Hennebelle2013} Hennebelle, P., \& Charbonnel, C. 2013, EAS Publications Series, 62
\bibitem[Hincelin et al. (2016)]{Hincelin2016} Hincelin, U., Commer\c{c}on, B., Wakelam, V., et al. 2016, ApJ, 822, 12
\bibitem[Hummer \& Rybicki (1971)]{Hummer1971}Hummer, D. G., \& Rybicki, G. B. 1971, MNRAS, 152(1), 1-19
\bibitem[Inutsuka et. al (2010)]{inu2010} Inutsuka, S. I., Machida, M. N., \& Matsumoto, T. 2010, ApJL, 718, L58
\bibitem[Larson (1969)]{Larson1969} Larson, R. B. 1969, MNRAS, 145, 271
\bibitem[Machida et al. (2006)]{Machida2006} Machida, M. N., Omukai, K., Matsumoto, T., \& Inutsuka, S. I. 2006, ApJL, 647, L1
\bibitem[Machida et al. (2011)]{Machida2011} Machida, M. N., Inutsuka, S. I., \& Matsumoto, T. 2011, PASJ, 63, 555
\bibitem[Masson et al. (2016)]{Masson2016} Masson, J., Chabrier, G., Hennebelle, P., Vaytet, N., Commer\c{c}on, B. 2016, A\&A, 587, A32
\bibitem[Masunaga, Miyama, \& Inutsuka(1998)]{inu98} Masunaga, H., Miyama, S., Inutsuka, S.~I. 1998 \apj, 495, 346
\bibitem[Masunaga, Miyama, \& Inutsuka(2000)]{inu2000} Masunaga, H., \& Inutsuka, S. I.  2000 \apj, 531(1), 350.
\bibitem[Mihalas \& Mihalas (1984)]{Mihalas1984} Mihalas, D., \& Mihalas, B. W. 1984, Foundations of Radiation Hydrodynamics (New York: Oxford Univ. Press)
\bibitem[Nordlund et al. (2014)]{Nordlund2014} Nordlund, \r{A}., Haugb\o lle, T., K\"uffmeier, M., et al. 2014, IAU Symposium, 299, 131
\bibitem[Seifried et al. (2012)]{Seifried2012} Seifried, D., Banerjee, R., Pudritz, R. E., \& Klessen, R. S. 2012, MNRAS, 423, L40
\bibitem[Seifried et al. (2016)]{Seifried2016} Seifried, D., S\'anchez-Monge, \'A.,Walch, S., \& Banerjee, R. 2016, MNRAS, 459, 1892
\bibitem[Spiegel(1957)]{TRM} Spiegel, E.~A. 1957 \apj, 126, 202
\bibitem[Stamatellos et al.(2007)]{Stamatellos2007} Stamatellos, D., Whitworth, A. P., Bisbas, T., \& Goodwin, S. 2007, A\&A, 475(1), 37-49
\bibitem[Stamatellos \& Whitworth (2008)]{Stamatellos2008} Stamatellos, D., \& Whitworth, A. P. 2008, MNRAS, 392(1), 413-427
\bibitem[Stone, Mihalas \& Norman (1992)]{Stone1992} Stone, J. M., Mihalas, D., \& Norman, M. L. 1992, ApJ Supplement Series, 80, 819-845
\bibitem[Tomida, Okuzumi, \& Machida (2015)]{Tomida2015} Tomida, K., Okuzumi, S., \& Machida, M. N. 2015, ApJ, 801, 117
\bibitem[Tomisaka (2002)]{Tomisaka2002} Tomisaka, K. 2002, \apj, 575, 306
\bibitem[Tscharnuter \& Winkler (1979)]{Tscharnuter1979} Tscharnuter, W. M., \& Winkler, K. H. 1979, Computer Physics Communications, 18(2), 171-199
\bibitem[Tsukamoto et al.(2015a)] {Tsukamoto2015a} Tsukamoto, Y., Iwasaki, K., Okuzumi, S., Machida, M. N., \& Inutsuka, S. I. 2015, MNRAS, 452, 278
\bibitem[Tsukamoto et al.(2015b)] {Tsukamoto2015b} Tsukamoto, Y., Iwasaki, K., Okuzumi, S., Machida, M. N., \& Inutsuka, S. (2015), ApJL, 810, L26
\bibitem[Vaytet \& Haugb\o lle (2017)]{Vaytet2017} Vaytet, N., \& Haugb\o lle, T. 2017, A\&A, 598, A116
\bibitem[Vaytet et al. (2011)] {Vaytet2011} Vaytet, N., Audit, E., Dubroca, B., \& Delahaye, F. 2011, J. Quant. Spectrosc. Radiat. Transfer, 112(8), 1323
\bibitem[Vaytet et al. (2012)] {Vaytet2012} Vaytet, N., Audit, E., Chabrier, G., Commer\c{c}on, B., \& Masson, J. 2012. A\&A, 543, A60
\bibitem[Vaytet et al. (2013)] {Vaytet2013}Vaytet, N., Chabrier, G., Audit, E., Commer\c{c}on, B., Masson, J., Ferguson, J., \& Delahaye, F. 2013. A\&A, 557, A90
\bibitem[Visser, Bergin, \& J\o rgensen (2015)]{Visser2015} Visser, R., Bergin, E., J\o rgensen, J. 2015, A\&A, 577, A102
\bibitem[Winkler \& Newman (1980)]{Winkler1980} Winkler, K.-H., \& Newman, M. J. 1980, \apj, 236, 201
\bibitem[Wurster, Price, \& Bate (2016)]{Wurster2016} Wurster, J., Price, D. J., \& Bate, M. R. 2016, MNRAS, 457, 1037
\bibitem[Yorke, Bodenheimer \& Laughlin (1993)]{Yorke1993} Yorke, H. W., Bodenheimer, P., \& Laughlin, G. 1993, \apj, 411, 274-284


\end{thebibliography}
\end{document}